\documentclass[a4paper,11pt]{article}
\pdfoutput=1 

\usepackage{jheppub} 

\usepackage[T1]{fontenc} 

\usepackage{epsfig}
\usepackage{graphicx}
 \usepackage{placeins}
\usepackage{mathrsfs}
\usepackage{amssymb}
\usepackage{float}
\usepackage{verbatim}
\usepackage{amsmath}
\usepackage[toc,page]{appendix}
\usepackage[colorlinks=true,linkcolor=blue]{hyperref}

\expandafter\ifx\csname package@font\endcsname\relax\else
 \expandafter\expandafter
 \expandafter\usepackage
 \expandafter\expandafter
 \expandafter{\csname package@font\endcsname}%
\fi

\title{\boldmath An update on the two singlet Dark Matter model}

\author[a]{Tanushree Basak,}
\author[b]{Baradhwaj Coleppa,}
\author[b]{Kousik Loho}

\affiliation[a]{Department of Physics, IISHLS, Indus University,\\Ahmedabad 382 115, India}
\affiliation[b]{Indian Institute of Technology Gandhinagar,\\Gandhinagar 382 355, India}

\emailAdd{tanushreebasak.gd@indusuni.ac.in}
\emailAdd{baradhwaj@iitgn.ac.in}
\emailAdd{kousik.loho@iitgn.ac.in}

\abstract{We revisit the two real singlet extension of the Standard Model with a $Z_2\times Z_2^\prime$ symmetry. One of the singlet 
scalars $S_2$, by virtue of an unbroken $Z_2^\prime$ symmetry, plays the role of a stable dark matter candidate. The other scalar $S_1$, with 
spontaneously broken $Z_2$-symmetry, mixes with the SM Higgs boson and acts as the scalar mediator. 
We analyze the model by putting in the entire set of theoretical and recent experimental constraints. 
The latest bounds from direct detection Xenon1T experiment severely restricts the allowed region of parameter 
space of couplings. To ensure the dark matter satisfies the relic abundance criterion, we rely on the Breit-Wigner 
enhanced annihilation cross-section. Further, we study the viability of explaining the observed gamma-ray excess in 
the galactic center in this model with a dark matter of mass in the $\sim 36-51$ GeV window and present our conclusions.
}

\arxivnumber{2105.09044}

\makeatletter
\gdef\@fpheader{}
\makeatother

\begin{document} 
\maketitle
\flushbottom


\section{Introduction}

The elusive dark matter, consisting of 24\% of the universe \cite{planck2018,wmap9}, finds no explanation in the 
Standard Model (SM) of particle physics. Although there are several evidences \cite{rotation_curve,bullet,Metcalf:2003sz,lensing}
for the existence of dark matter (DM) on 
cosmological scales, direct evidences such as scattering of DM particles off nuclei are still lacking. 
The properties and interaction of dark matter are still unknown, except that it must be 
electrically neutral and stable on cosmic time scales. 
Therefore, the quest for identifying dark matter continues to motivate extensions beyond the SM. A 
popular choice for DM is the weakly interacting massive particle (WIMP) \cite{Bertone:2004pz,Bergstrom:2009ib,Arcadi:2017kky} scenario with the DM mass ranging from a few GeV to a few hundred GeVs. It is imperative that the WIMPs do not have sizable couplings to the SM sector lest they decay. A discrete symmetry under which the dark matter and the SM behave differently is a convenient mechanism with which to prohibit such decays. There is a large literature of WIMP candidates from various well-motivated Beyond the SM (BSM) scenarios. For example, in this context supersymmetry (SUSY),  neutralinos are widely studied in the literature  \cite{Jungman:1995df} as possible WIMPs. Neutralinos being the lightest supersymmetric particles (LSP) are odd under R-parity enabling them to be stable DM candidates. But
it is quite challenging to achieve a light dark matter below 100 GeV in minimal SUSY model owing to various 
phenomenological and experimental constraints \cite{Belanger:2012jn,Belanger:2013pna,Hagiwara:2013qya,Cao:2015efs,Chakraborti:2017dpu}.

In the context of minimal non-supersymmetric scenarios, the real singlet extension of the SM has been extensively studied 
\cite{McDonald:1993ex,Burgess:2000yq,Guo:2010hq,Bandyopadhyay:2010cc,Cline:2013gha,Beck:2021xsv,He:2008qm,He:2009yd}, where an imposed $Z_2$-symmetry ensures the stability of the extra singlet 
 rendering it a viable dark matter candidate. In such models, it has been 
observed that only a dark matter with mass nearly half that of the SM Higgs boson can survive the relic abundance constraints, which makes light DM with mass less than 60 GeV or so  
not viable even if they are consistent with direct detection bounds. A possible way out of this leads one to the next to minimal approach: a two singlet extension of the SM 
 with an added $Z_2\times Z_2^\prime$ discrete symmetry(for both singlets charged under same $Z_2$ see ref.\cite{Ghorbani:2014gka}). 
Previous studies on two singlet model in the pre-LHC era \cite{Abada:2011qb,Abada:2012hf} have shown that a DM with mass $\sim 10-20$ GeV and a light mediator is consistent 
with the CDMSII\cite{Ahmed:2008eu} and XENON100\cite{Aprile:2010um} bounds. Ref.\cite{Modak:2013jya}, considers similar models but in the context 
of slightly complicated multi component DM while Ref.\cite{Ahriche:2013vqa} focusses on Higgs phenomenology 
in the two singlet model. In a recent study \cite{Arhrib:2018eex}, the authors have taken into account self interactions of the DM 
and with a light mediator to explain the observed density profiles of dwarf galaxies and galaxies of the size of our Milky way.

The paper is organized as follows: in Sec.~\ref{sec:model}, we review the two-singlet model \cite{Abada:2011qb,Arhrib:2018eex} 
 with a $Z_2\times Z_2^\prime$ symmetry and compute the relevant couplings and mixings.  Then in Sec.~\ref{sec:constraints}, we study the implications of imposing various constraints on the model from collider searches and the vacuum stability 
bound. The invisible decay width of SM Higgs boson restricts the choice of certain parameters like the dark matter self coupling 
and its coupling to the SM Higgs. Sec.~\ref{sec:pheno} deals with the detailed dark matter phenomenology - we demonstrate in Sec.~\ref{sec:dd} that 
in order to be consistent with recent Xenon1T exclusion limits, the parameter space for couplings are quite tightly
constrained. Relic abundance (Sec.~\ref{sec:relic}) criteria are found to be satisfied only near the regions where the mass of the 
DM is almost half the mass of either the SM Higgs boson ($h$) or the other scalar ($H$) where the 
Breit-Wigner enhancement is significant - elsewhere the annihilation cross-section is too low leading to over-abundance. In Sec.~\ref{sec:gre}, we discuss the reported gamma ray excess observed in the galactic center which 
has grabbed a lot of attention lately. Choosing the DM mass in the range  $\sim 36-51$ GeV with $m_{H}$ ranging between $70-104$ GeV, we demonstrate that the
galactic center gamma ray excess \cite{Goodenough:2009gk,Boyarsky:2010dr,Hooper:2010mq,Abazajian:2012pn,Gordon:2013vta} can be accommodated
in this model. This again is because the annihilation cross-section of DM-DM into $b\bar b$ is enhanced by virtue of the scalar resonance.  Finally in Sec.~\ref{sec:conclusion}, we conclude with remarks on the future scope of this model.


\section{The Two-singlet Model}
\label{sec:model}

One of the simplest ways to incorporate a dark matter candidate is by extending the scalar sector of the SM. In this paper, we concentrate on the extension of the SM with two real scalars $S_1$ and $S_2$ with an imposed $Z_2\times Z_2^\prime$ symmetry. Here we capture the essential elements of the model pertaining to the symmetry breaking pattern - more details about the model can be found in Refs.~\cite{Abada:2011qb,Arhrib:2018eex,Hamada:2020wjh}. 

The model is constructed such that all SM particles are even under both the $Z_2$'s while the extra scalar singlets transform in the following manner:
\begin{align}
\begin{split}
&S_1\xrightarrow{Z_2}-S_1,\, S_1\xrightarrow{Z_2^\prime}S_1, \\
&S_2\xrightarrow{Z_2}S_2,\,S_2\xrightarrow{Z_2^\prime}-S_2.
\end{split}
\end{align}
Both $S_1$ and $S_2$ interact with the SM particles solely through a Higgs-portal. The $Z_2$ symmetry is broken spontaneously while the $Z_2^\prime$ remains unbroken which continues to insure the stability of $S_2$, thus making it a potential dark matter candidate (we will consider the WIMP scenario in this paper). As we will see below, as a consequence of the Higgs-portal structure, interactions of the DM with the SM particles arise purely from the DM-Higgs boson couplings. We begin by writing down the  Lagrangian of the scalar sector: 

\begin{equation}\label{eq:new-scalar_L}
\mathcal{L}_s=( D^{\mu} \Phi )^{\dagger} D_{\mu}\Phi + \frac{1}{2} \partial^{\mu} S_1 \partial_{\mu}S_1 +
\frac{1}{2} \partial^{\mu} S_2 \partial_{\mu}S_2 - V(\Phi,S_1,S_2 ), \,
\end{equation}
where $\Phi$ is the usual SM Higgs doublet. The scalar potential admits all terms respecting the gauge and the discrete $Z_2\times Z_2^\prime$ symmetries:
\begin{equation}\label{eq:scalar_potential}
\begin{split}
V=&-\frac{m_0^2}{2}(\Phi^\dagger\Phi)-\frac{m_1^2}{2}S_1^2+\frac{m_2^2}{2}S_2^2+\frac{\lambda_0}{4}(\Phi^\dagger\Phi)^2+\frac{\lambda_1}{4}S_1^4+\frac{\lambda_2}{4}S_2^4+\lambda_{01}(\Phi^\dagger\Phi)S_1^2\\
&+\lambda_{02}(\Phi^\dagger\Phi)S_2^2+\lambda_{12}S_1^2S_2^2.
\end{split}
\end{equation}
The scalar doublet $\Phi$ develops a vacuum expectation value (vev) $v_0$ breaking the $SU(2)_L\times U(1)_Y$ spontaneously. Denoting the vev of $S_1$ by $v_1$, the two fields can be written in unitary gauge after symmetry breaking in the form
\begin{equation}\label{eq:SSB}
\Phi =\frac{1}{\sqrt{2}}  
\begin{pmatrix}
0\\
v_0+h_0\\
\end{pmatrix}; 
	\hspace{0.5cm} S_1={v_1+h_1}\, ,
\end{equation} 
where both $v_0$ and $v_1$ are real and positive and $h_0$ and $h_1$ are excitations around the respective vevs. The Domain Wall problem \cite{Zeldovich:1974uw} can be evaded by explicitly breaking the $Z_2$ symmetry corresponding to $S_1$ by introducing in the Lagrangian a term linear in $S_1$ \cite{Arhrib:2018eex}. Assuming the corresponding coupling with cubic mass dimension to be Planck mass suppressed, we have not explicitly shown such a term in Eqn.~\ref{eq:scalar_potential} as it becomes important only at energies close to the Planck scale. Extremization of the potential gives the following relations between the different parameters:
\begin{equation}\label{eq:extremisation}
m_0^2=\frac{\lambda_0v_0^2}{2}+2\lambda_{01}v_1^2;
\hspace{0.1in}
m_1^2=\lambda_1v_1^2+\lambda_{01}v_0^2.
\end{equation}

After symmetry breaking, the mass matrix for the scalar sector can be written as 
\begin{equation}\label{eq:mass_matrix}
\mathcal{M}_\textrm{s}=\begin{pmatrix}
h_0&h_1\\
\end{pmatrix} 
\begin{pmatrix}
\frac{\lambda_0v_0^2}{4}&\lambda_{01}v_0v_1\\
\lambda_{01}v_0v_1&\lambda_1v_1^2\\
\end{pmatrix} 
\begin{pmatrix}
h_0\\
h_1\\
\end{pmatrix} +\frac{1}{2}(m_2^2+\lambda_{02}v_0^2+2\lambda_{12}v_1^2)S_2^2.
\end{equation}
The off-diagonal terms indicate mixing - the mass matrix can be diagonalized by a unitary rotation in the usual manner and the mass eigenstates can be written as

\begin{equation}\label{eq:eigenstates}
\begin{pmatrix}
H\\
h\\
\end{pmatrix} =\begin{pmatrix}
\cos\alpha&\sin\alpha\\
-\sin\alpha&\cos\alpha\\
\end{pmatrix} \begin{pmatrix}
h_0\\
h_1\\
\end{pmatrix},
\end{equation}
where $h$ and $H$ are the two mass eigenstates and the mixing angle $\alpha$ is given by 
\begin{equation}\label{eq:mixing_angle}
\tan2\alpha=\frac{2\lambda_{01}v_0v_1}{\frac{\lambda_0v_0^2}{4}-\lambda_1v_1^2}.
\end{equation}
While one could directly write down the mass eigenvalues $m_{h,H}$ from Eqn.~\ref{eq:mass_matrix}, for our purposes it is more useful to use these as inputs so we can use them to re-express other Lagrangian parameters. In particular, the quartic self-couplings $\lambda_0$ and $\lambda_1$ can be written as

\begin{align}
\begin{split}
\lambda_0&=\frac{2}{v_0^2}\bigg(\frac{m_h^2+m_H^2}{2}+\sqrt{\frac{(m_h^2-m_H^2)^2}{4}-4\lambda_{01}^2v_0^2v_1^2}\bigg), \\
\lambda_1&=\frac{1}{2v_1^2}\bigg(\frac{m_h^2+m_H^2}{2}-\sqrt{\frac{(m_h^2-m_H^2)^2}{4}-4\lambda_{01}^2v_0^2v_1^2}\bigg).
\end{split}
\label{eq:lambdas}
\end{align}
We will use these formulas extensively while analyzing the parameter space of this model in what follows. Finally, rewriting the Lagrangian in the mass basis, we can collect the cubic and quartic interaction terms of the scalars:
\begin{align}
\begin{split}
V_{\textrm{cubic}}=&\lambda_{HHH}H^3+\lambda_{hhh}h^3+\lambda_{HHh}H^2h+\lambda_{Hhh}Hh^2+\lambda_{HS_2S_2}HS_2^2+\lambda_{hS_2S_2}hS_2^2, \\
V_{\textrm{quartic}}=&\lambda_{HHHH}H^4+\lambda_{hhhh}h^4+\lambda_{S_2S_2S_2S_2}S_2^4+\lambda_{HHhh}H^2h^2+\lambda_{HHS_2S_2}H^2S_2^2\\
&+\lambda_{hhS_2S_2}h^2S_2^2
+\lambda_{HHHh}H^3h+\lambda_{Hhhh}Hh^3+\lambda_{HhS_2S_2}HhS_2^2.
\end{split}
\label{eq:interactions}
\end{align}
The explicit form of the various couplings in the expressions above are given in detail in Appendix \ref{sec:appendix}.


\section{Constraints on the Two-singlet model}
\label{sec:constraints}
In this section, we will impose all theoretical and experimental constraints upon the parameter space of the model to figure out the surviving areas. Before we begin, we note that we will need to fix the values of certain parameters 
from either experimental observations or from phenomenological requirements: specifically, we fix $v_0=246$ GeV, the scale of Electroweak Symmetry Breaking (EWSB) and one of the two Higgses in the model to be the observed $126$ GeV Higgs boson. We do the analysis for two choices: fixing the lighter of the two to be the SM Higgs and choosing the other to have a mass $m_H=250$ GeV, and choosing the heavier one to be the SM Higgs fixing the lighter Higgs at $m_h=80$ GeV. For both these cases, we will analyze and scan the parameter space of the couplings ($\lambda_{01}$, $\lambda_{02}$, $\lambda_{12}$) and put bounds on their values from various constraints. $\lambda_{01}$ and $\lambda_{02}$ are the Higgs portal couplings while the $\lambda_{12}$ represents the DM interaction with the other scalar $S_1$. While it would certainly be more useful to directly constrain the physical couplings as given in Appendix \ref{sec:appendix}, as can be seen from that table, the expressions can be rather unwieldy and hence in this paper we restrict ourselves to analyzing the Lagrangian parameters instead. We begin with the vacuum stability conditions.

\subsection{Vacuum Stability}

\begin{figure}[t!]
\centering
\includegraphics[scale=0.59]{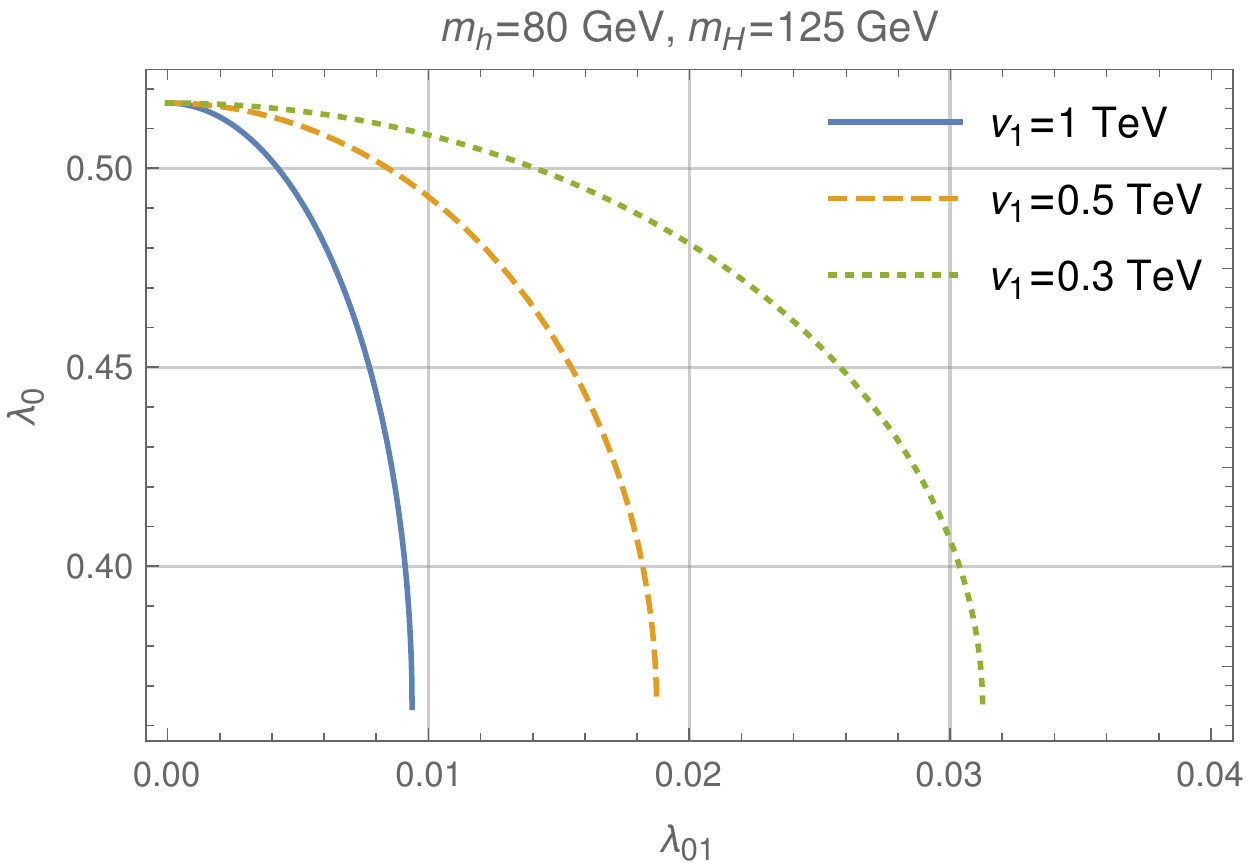}
\includegraphics[scale=0.59]{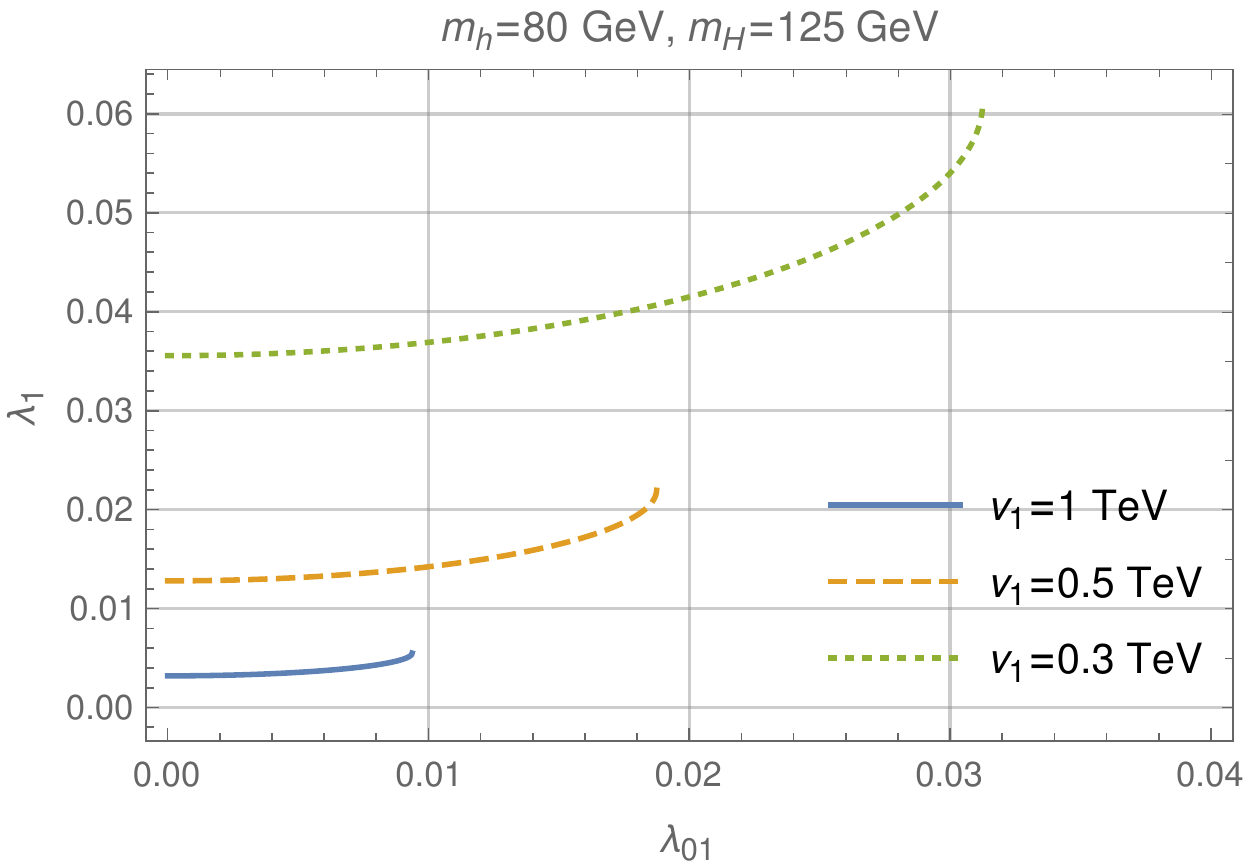}\\
\includegraphics[scale=0.59]{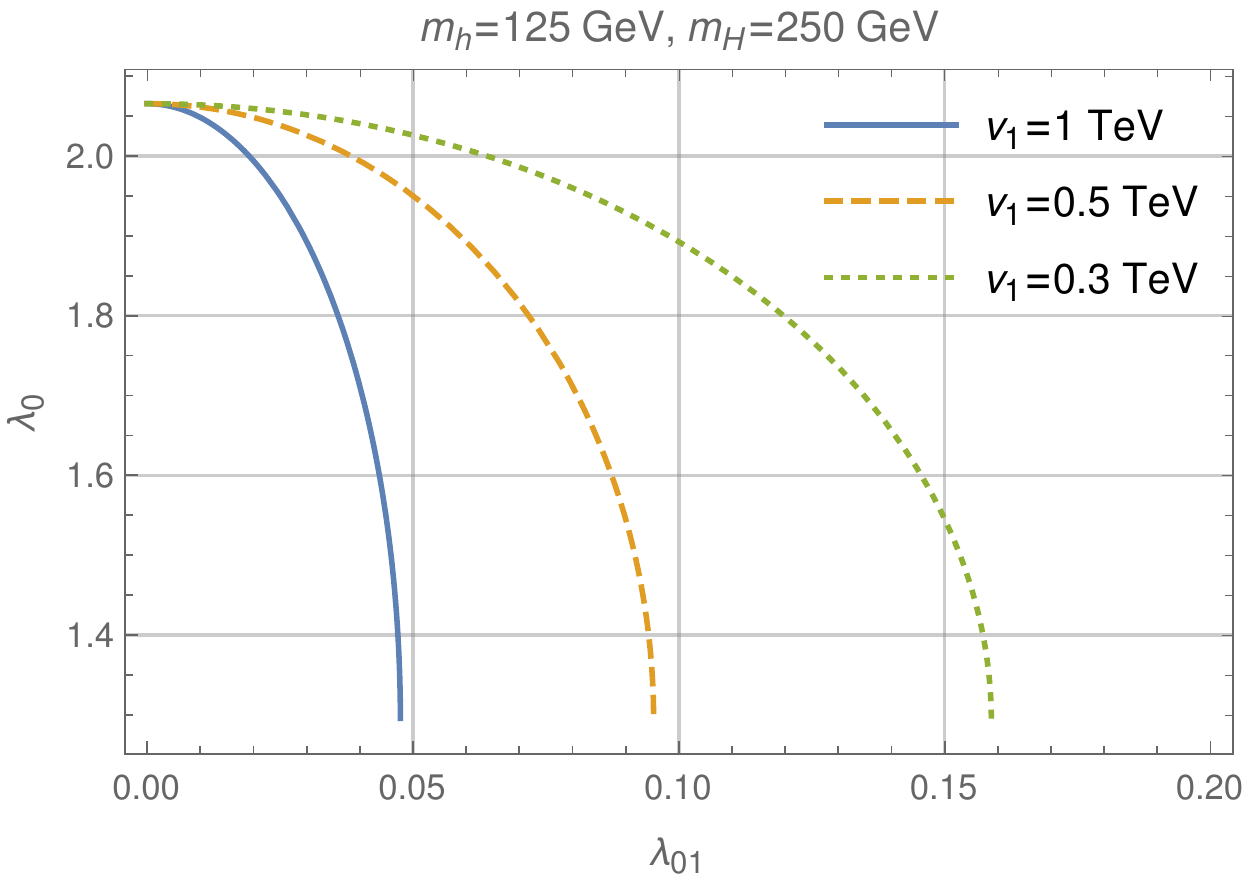}
\includegraphics[scale=0.59]{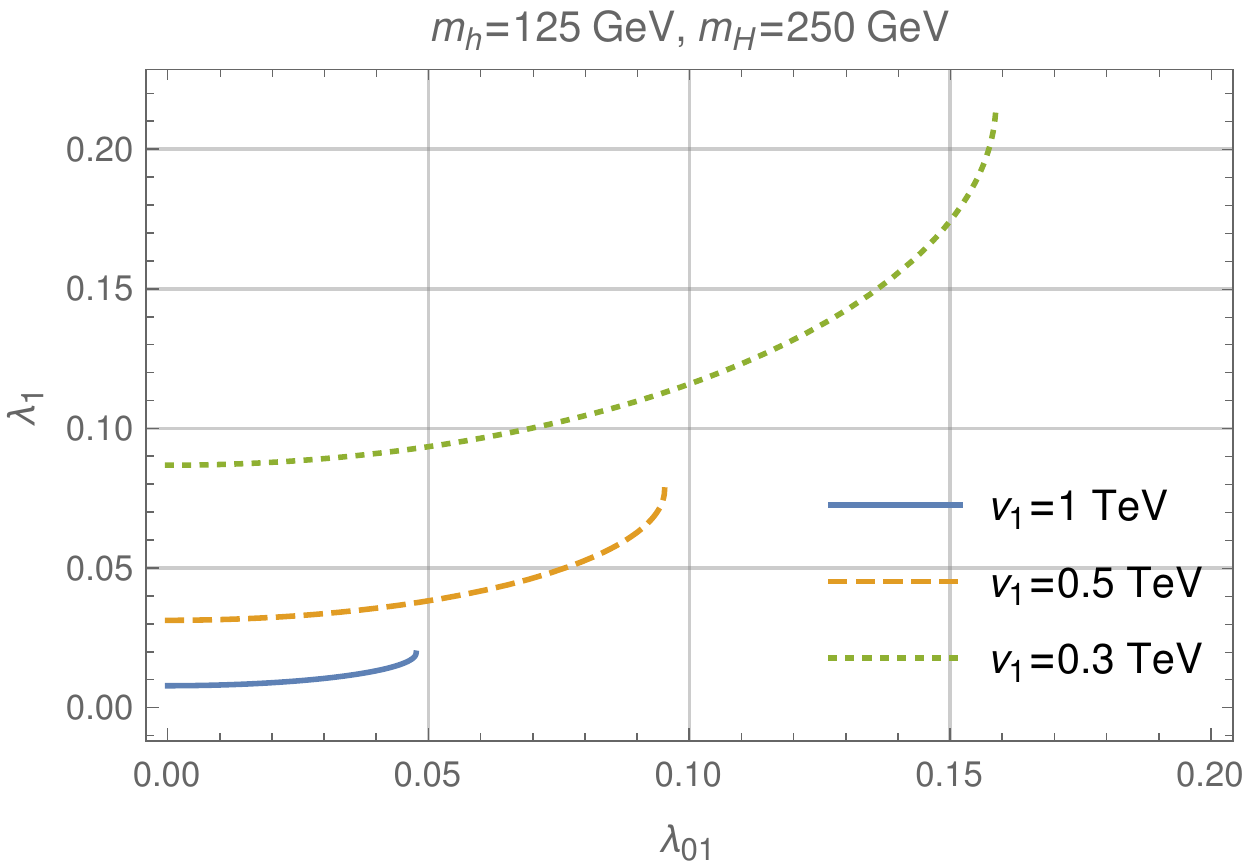}
\caption{Bound on $\lambda_{01}$ (top left) due to positivity constraints on $\lambda_0$ for various values of $v_1$: 
1 TeV (solid), 0.5 TeV (dashed) and 0.3 TeV (dotted) and that on  $\lambda_{1}$ (top right) for the $m_H=125$ GeV case 
for the same set of $v_1$ values (colors online). The bottom panel shows the bounds on these parameters for the $m_h=125$ 
GeV case. Note that these are \emph{not} contour plots - only those points displayed on each curve is allowed and the 
rest of the parameter space is unphysical.}
\label{fig:vs1}
\end{figure}

Calculating the vacuum stability conditions analytically for any model is in general quite involved. However, 
in lieu of doing a detailed analysis \cite{Modak:2013jya,Ghosh:2017fmr}, we use a graphical analysis to get an idea of the upper 
bound on $\lambda_{01}$. An important criteria for SSB is the `wrong' sign of the mass term: there has to be a relative sign difference between the mass term and the term 
containing the fourth power of the scalar field in the Lagrangian. Thus ensuring that both $\Phi$ and $S_1$ engineer the breaking of $SU(2)\times U(1)$ down to $U(1)_{\textrm{em}}$ demands 
that the parameters $\lambda_0$ and $\lambda_1$ be real and positive. With the chosen benchmark values of 
$m_h$, $m_H$, and $v_0$, and using Eqns.~\ref{eq:lambdas}, we display the extent of $\lambda_{01}$ values consistent with the positivity condition of $\lambda_0$ and $\lambda_1$ 
for various values of $v_1$  in Fig.~\ref{fig:vs1}. To understand the nature of these plots, we first note from Eqns.~\ref{eq:lambdas} that in the limit $\lambda_{01}\to 0$, $\lambda_0\approx 2m_h^2/v_0^2$ and $\lambda_1\approx m_H^2/2v_1^2$. For the chosen benchmark points, we expect $\lambda_1$ to be more constrained than $\lambda_0$ for a given $\lambda_{01}$, particularly for larger $v_1$ values. Looking at the $\lambda_0-\lambda_{01}$ parameter space for instance, we see that for $v_1=1$ TeV, for a $\lambda_{01}$ value of around $\sim$ 0.008 for the $m_H=125$ GeV case $\lambda_0$ can reach a value of around $0.4$, while the upper limit on $\lambda_1$ for the same $\lambda_{01}$ is around 0.004. This trend is consistent with the $m_h=125$ GeV case as well as can be readily seen in Fig.~\ref{fig:vs1}. In fact, turning this argument around, we see that the parameter $\lambda_{01}$ is itself highly constrained even from this simple analysis - if one forces the value of $\lambda_{01}$ higher, the couplings $\lambda_0$ (or $\lambda_1$) quickly become imaginary thus affecting the symmetry breaking structure of the model. For the purposes of our analysis, conclusions drawn on $\lambda_{01}$ are highly relevant as this higgs portal coupling (see Eqn.~\ref{eq:scalar_potential}) will play a crucial role in dark matter phenomenology. We note, however, that the couplings $\lambda_{02}$ and $\lambda_{12}$ are unaffected by this analysis as $S_2$ does not mix with the other scalars in the model.\footnote{Note that we have considered   $\lambda_0$ values $>1$ in the $m_h=125$ GeV case - however this is only a Lagrangian parameter and \emph{not} a physical coupling.}


\subsection{Collider Constraints}
\label{sec:collider}

A major collider constraint on Higgs-portal dark matter sector comes from the Branching Ratio (BR) of the Higgs invisible decays:
$$\textrm{BR}_{h\rightarrow inv}=\frac{\Gamma_{h\rightarrow inv}}{\Gamma_{h\rightarrow SM}+\Gamma_{h\rightarrow inv}}.$$
Given the presence of extra scalars that couple to the SM Higgs, this constraint becomes an important one for the present model. ATLAS \cite{ATLAS-CONF-2020-052} and CMS \cite{Sirunyan:2018owy} have provided experimental limits on the branching ratio of invisible Higgs decays - the bound from ATLAS is the strongest till date with an observed value of around 11\%. To translate this bound to the present model, we need to first fix the masses. In anticipation of future results (laid out in Secs.~\ref{sec:relic} and \ref{sec:gre}), we choose $m_{S_2}=40$ GeV. In the $m_H=125$ GeV case with both $h$ and $S_2$ lighter, both decays $H\to hh$ and $H\to S_2S_2$ need to be considered - however, for the chosen benchmark point ($m_h=80$ GeV), the only tree-level decay that is relevant is the latter. For the other case ($m_h=125$ GeV), only the decay $h\to S_2S_2$ is relevant as we have chosen the mass of the $H$ to be higher than that of the $h$. Thus, the relevant couplings in these two cases are $\lambda_{HS_2S_2}$ and $\lambda_{hS_2S_2}$. Referring to Appendix \ref{sec:appendix}, we see that these couplings involve the four parameters\footnote{The couplings also involve $v_0$ which is fixed at 246 GeV.} $\lambda_{02}$, $\lambda_{12},$ the mixing angle $\alpha$, and $v_1$. Of these, with the use of Eqns.~\ref{eq:mixing_angle} and \ref{eq:lambdas}, the parameter $\alpha$ can be traded for the masses, $\lambda_{01}$, and $v_1$.  We choose the lowest admissible value of $\lambda_{01}$ (0.008 for the $m_H=125$ GeV case and 0.04 for the $m_h=125$ GeV case) from Fig.~\ref{fig:vs1}. While other values of $\lambda_{01}$ could certainly be considered, the numerical difference in this coupling (as seen from Fig.~\ref{fig:vs1}) is not very large, and the chosen $\lambda_{01}$ value is consistent for all three choices of $v_1$ that we consider here.  

\begin{figure}[t!]
\centering
\includegraphics[scale=0.5]{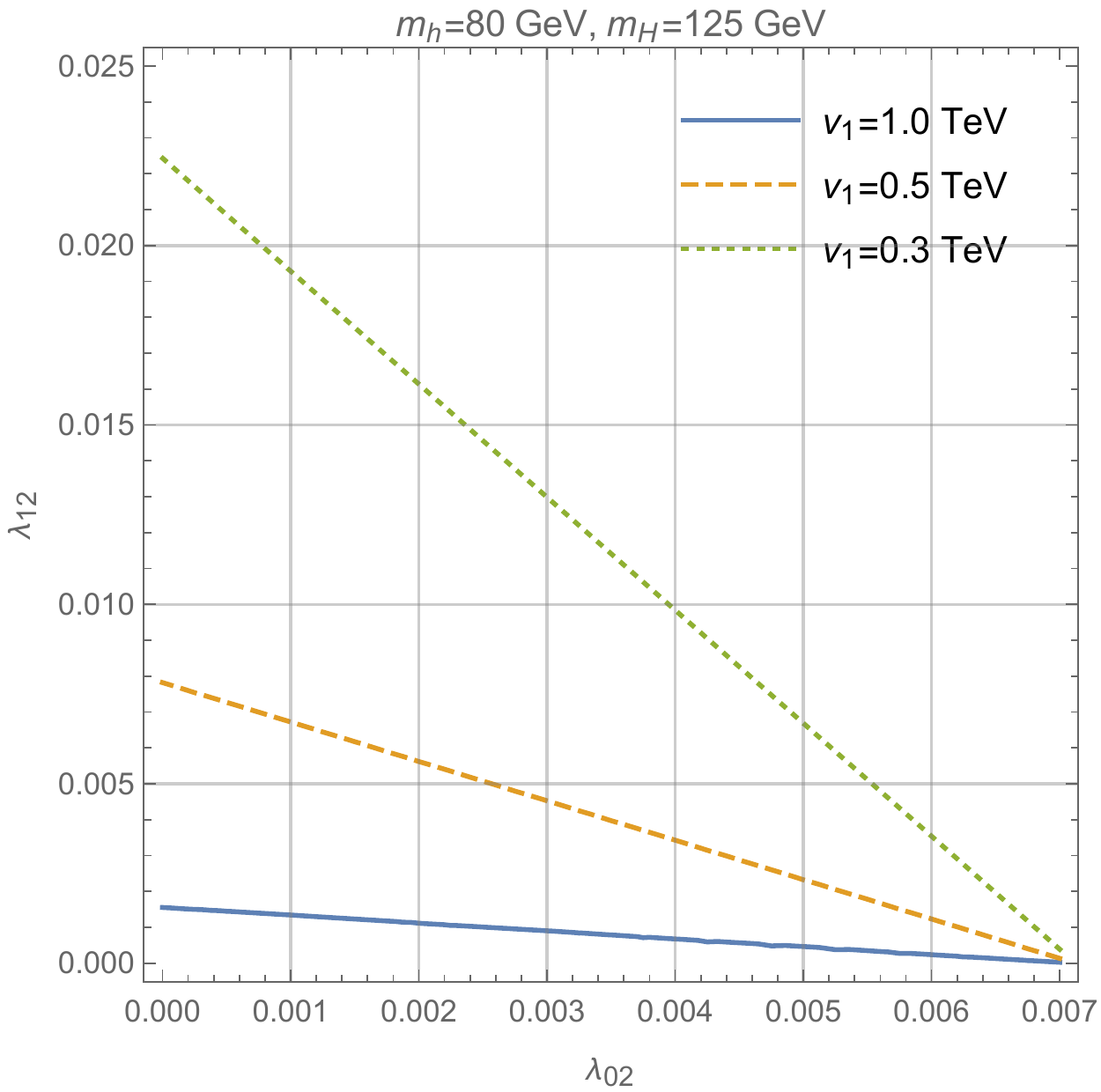}
\includegraphics[scale=0.5]{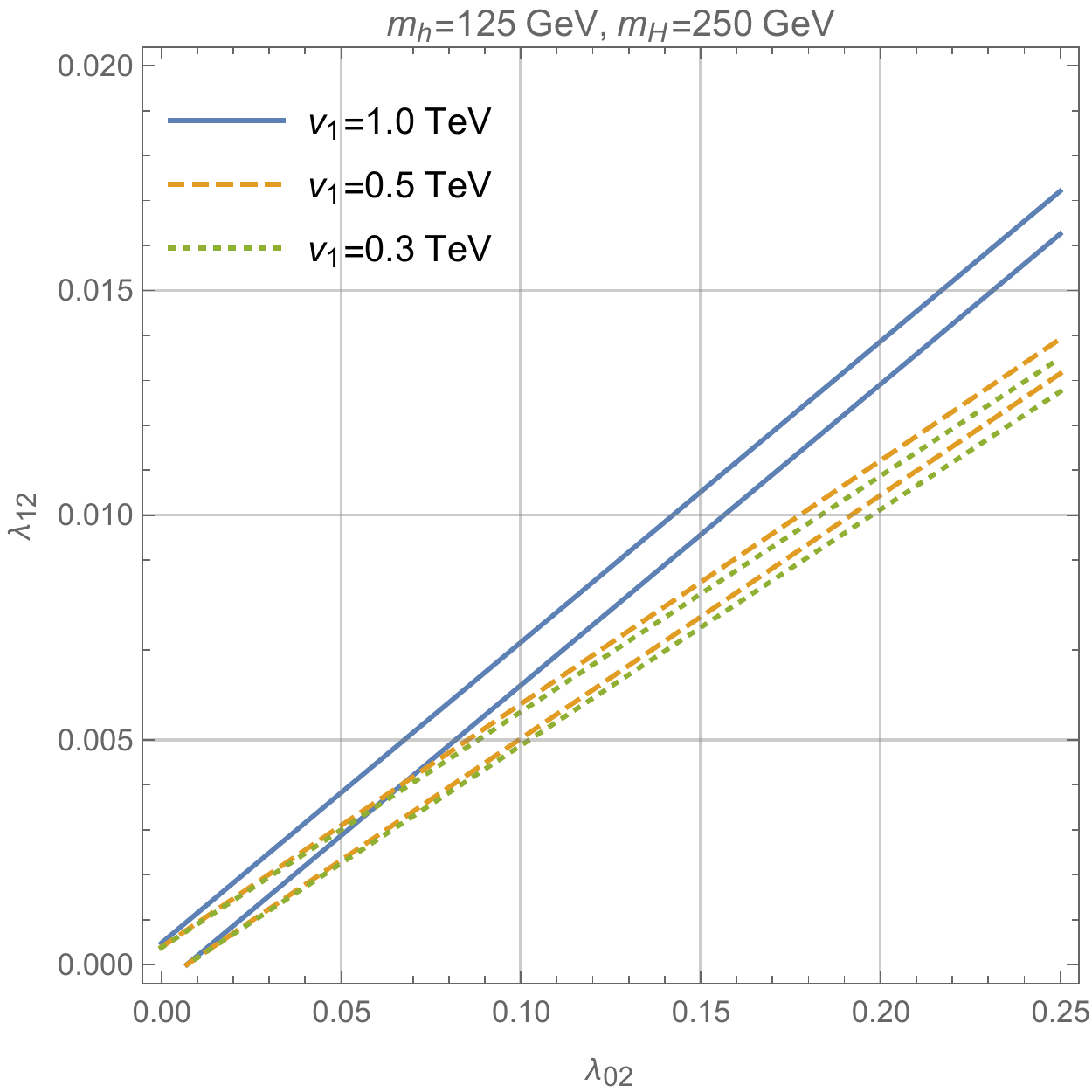}
\caption{Contours of BR$=$11\% for $v_1=$ 1 TeV (solid), 0.5 TeV (dashed) and 0.3 TeV (dotted) for the $m_H=125$ GeV case (left) and the $m_h=125$ GeV case (right) (colors online). The allowed region for the $m_h=125$ GeV case on the right is the band between the corresponding lines.}
\label{fig:invisible}
\end{figure}

In Fig.~\ref{fig:invisible}, we illustrate the invisible decay constraints in this model by displaying contours of  11\% BR\footnote{To calculate these BRs, we considered the dominant decays of the higgs: $b\bar{b}$, $\tau\bar{\tau}$, and $S_2S_2$, and neglected off-shell decays like $WW$ and $ZZ$.} in the $\lambda_{02}-\lambda_{12}$ parameter space for three different values of $v_1$ for both the $m_H=125$ GeV (left) and $m_h=125$ GeV (right) cases. In the former case, it is seen that lowering the scale $v_1$ increases the range of allowed values of $\lambda_{12}$ while in the latter it has the opposite effect - in this case it is seen that lowering $v_1$ shifts the allowed band toward a more restricted range for $\lambda_{12}$. In general, we see that the $m_H=125$ GeV case prefers larger $\lambda_{12}$ and slightly smaller $\lambda_{02}$ values compared to the $m_h=125$ GeV case. We see from the nature of the couplings (see Appendix \ref{sec:appendix})
\begin{align}
\begin{split}
\lambda_{HS_2S_2}&=\lambda_{02}v_0\cos\alpha+2\lambda_{12}v_1\sin\alpha \\
\lambda_{hS_2S_2}&=-\lambda_{02}v_0\sin\alpha+2\lambda_{12}v_1\cos\alpha
\end{split}
\label{eq:higgs_cubic}
\end{align}
that for a given $\alpha$, the relative importance of the $\lambda_{02}$ and $\lambda_{12}$ pieces will be different in the two cases.

 A couple of general features arise from these two constraints put together:
 \begin{itemize}
 \item In spite of being a rather minimal extension of the SM, this model has quite a few independent parameters. We observe that while lowering $v_1$ shifts the allowed regions in the parameter space, the numerical values of the couplings do not change significantly for the different choices of $v_1$ employed. Thus, we will fix $v_1=1$ TeV henceforth.
 \item Fixing $\lambda_{01}=0.008$ (0.04) for $m_H=80$ GeV (250 GeV) and looking at the $\lambda_{02}-\lambda_{12}$ parameter space, we conclude that all these couplings are roughly constrained to lie in the numerical range $10^{-3}-10^{-2}$.
 \end{itemize}

\subsection{LHC Higgs searches}
\label{sec:collider_direct}

\begin{figure}[t!]
\centering
\includegraphics[scale=0.5]{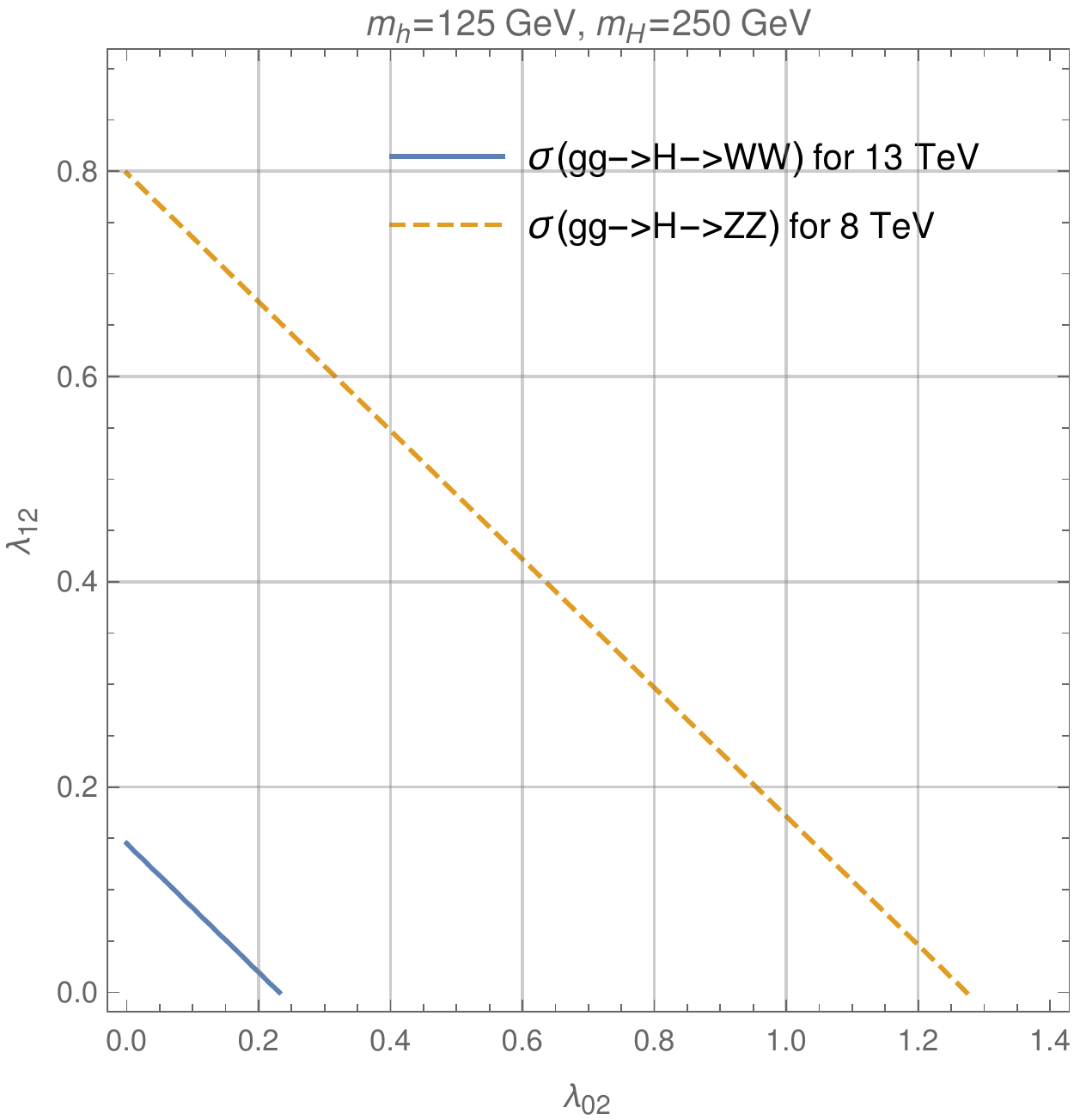}
\includegraphics[scale=0.51]{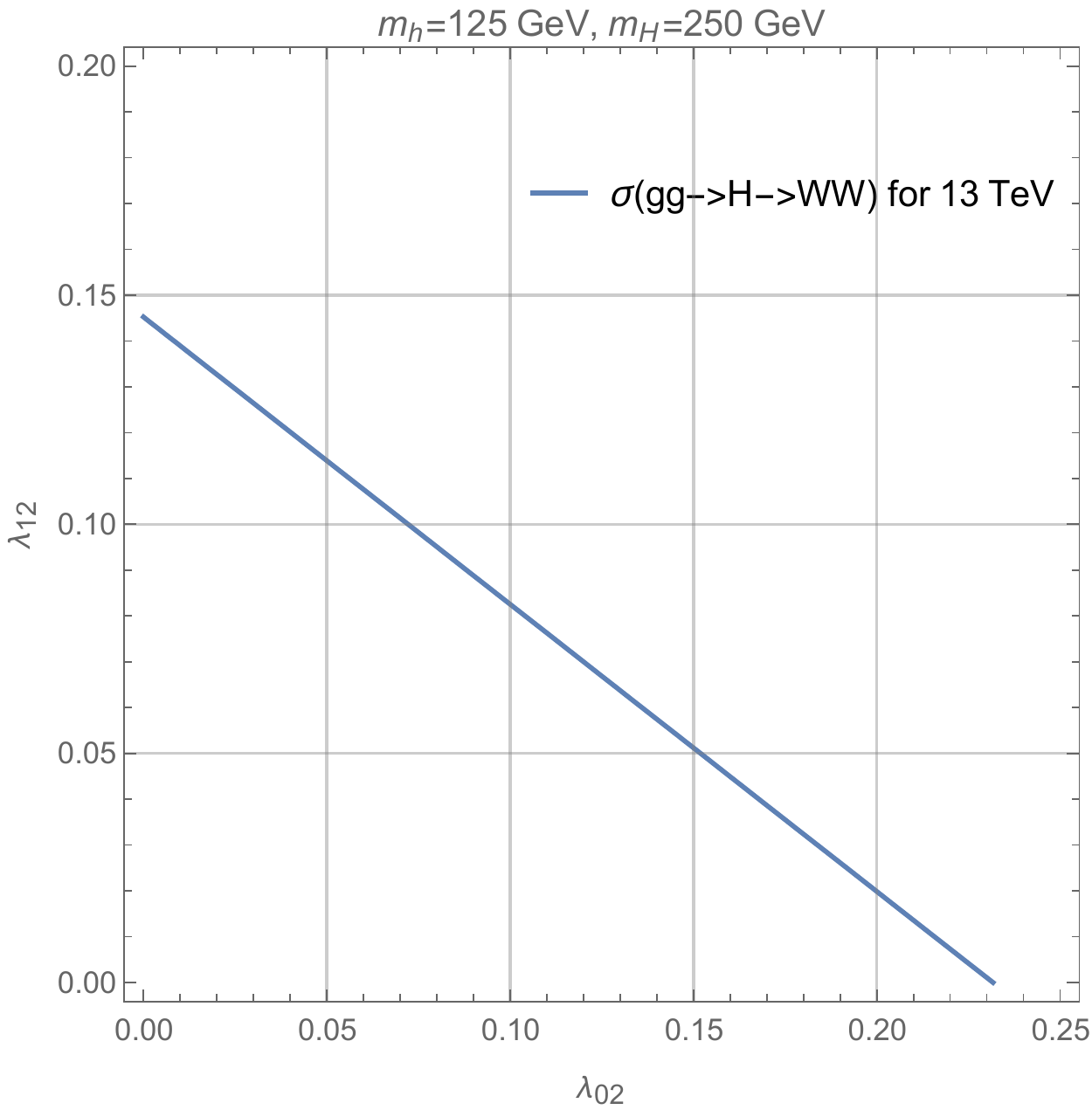}
\caption{Allowed regions in the $\lambda_{02}-\lambda_{12}$ parameter space from searches for the heavy scalar $H$ at the LHC in the diboson channel $gg\to H \to WW/ZZ$ with the gauge bosons decaying either leptonically or hadronically (left) and a magnified view of the 13 TeV constraints (right) .}
\label{fig:collider}
\end{figure}

After the discovery of the SM-like Higgs, the LHC experiments have continued to look for heavier scalars in a variety of final states. Non-observation of such particles constrains the parameter space of theoretical models that admit such states in their spectrum, and thus we need to understand how this affects the present model. For the first benchmark point ($m_h=80$ GeV, $m_H=125$ GeV), there are not many direct searches that impact the model. However, for the second choice ($m_h=125$ GeV, $m_H=250$ GeV), searches for a heavy scalar in the channel $gg\to H \to WW/ZZ$ provide useful constraints, with the gauge bosons further decaying to various hadronic and leptonic states whose combined effect \cite{Sirunyan:2019pqw,Aad:2015kna} has been taken into account. In translating this bound, we should also account for the fact that the $HWW$ coupling is scaled by a factor of $\cos\alpha$ relative to the SM Higgs coupling - however, like in the previous section we trade the parameter $\alpha$ for $\lambda_{01}$ and $v_1$, choosing $\lambda_{01}=0.04$ as before. In Fig.~\ref{fig:collider}, we present the direct bounds from the LHC searches in the $\lambda_{02}-\lambda_{12}$ parameter space from both the 8 TeV and the 13 TeV data. Understandably, the 13 TeV bounds are tighter - however, we note that the Higgs invisible decay limits (Fig.~\ref{fig:invisible}) constrain this parameter space much more than the results of the direct search experiments.

\FloatBarrier

\section{Dark matter phenomenology in the Two-singlet Model}
\label{sec:pheno}

The two-singlet extension of the scalar sector of the SM has been studied in the literature from the dark matter point of view for different symmetry breaking patterns of the discrete $Z_2\times Z'_2$. Our work complements these and aims to address the results of the latest experimental efforts. The direct detection constraints being comparatively less restrictive for lighter ($<10$ GeV) dark matter masses can give rise to very exciting phenomenology \cite{Maniatis:2020ois}. Investigations \cite{Abada:2011qb,Ahriche:2013vqa,Abada:2012hf} have been carried out in good detail to restrict this model with various constraints from collider and direct searches - however some of these bounds are presently outdated. The two-singlet extension scenarios have also been explored in some other contexts such as the exchange-driven freeze out mechanism \cite{Maity:2019hre}, dwarf galaxy density profiles \cite{Arhrib:2018eex}, MeV scale SIMP dark matter with a $3\rightarrow2$ process dictating the relic \cite{Mohanty:2019drv}, and mitigating direct detection bounds under a mix of annihilation and co-annihilation of multicomponent dark matter \cite{Bhattacharya:2017fid}. In this work, we update the exclusion limits from the latest collider and direct search results with a focus on relaxing the direct detection constraints taking advantage of the destructive interference between the two higgs exchange diagrams and its implications for possible future results. In addition, we also address the gamma-ray excess observation from the galactic center \cite{Goodenough:2009gk,Boyarsky:2010dr,Hooper:2010mq,Abazajian:2012pn,Gordon:2013vta} and Fermi bubble - while this been explored in the light of a multicomponent dark matter scenario (see for example \cite{Modak:2013jya}), we demonstrate here that two-singlet extension models can also account for this excess reasonably well taking advantage of the Breit-Wigner resonance.


\subsection{Direct detection}
\label{sec:dd}

The direct detection experiments probe the cross-section for the scattering of the DM particle off a nucleon at non-relativistic limits. 
The effective lagrangian for nucleon-dark matter interaction is given by
\begin{equation}\label{eq:eff_lag}
 L_{eff}=a_N N\bar{N}S_2^2,
\end{equation}
where $a_N$ is the effective coupling constant between the DM and the nucleon. The spin-independent scattering cross-section for the scattering of the DM off a proton or neutron goes as
$\sigma_{p,n}^{SI}\sim\mu^2f_{p,n}^2$ \cite{Ellis:2000ds}, where $\mu$ is the reduced mass and  $f_{p,n}$ is the 
hadronic matrix element:
\begin{equation}\label{eq:scalarterms}
f_{p,n} = \sum_{q=u,d,s}  f_{Tq}^{(p,n)} a_q  \frac{m_{p,n}}{m_q}  + \frac{2}{27}f_{TG}^{(p,n)}
\sum_{q=c,b,t} a_q \frac{m_{p,n}}{m_q}, \nonumber
\end{equation}
where $a_q$ is the (model-dependent) coupling between the DM and quarks and the quantities $f^{(p,n)}_{Tq}$ and $f^{(p,n)}_{TG}$ \cite{Shifman:1978zn} 
are defined as
\begin{equation}
f^p_{Tq}m_p\langle p|p\rangle\equiv\langle p|m_q\bar{q}q|p\rangle; \,\,f_{TG}^{(p,n)} = 1 -\sum_{q=u,d,s}f_{Tq}^{(p,n)},
\end{equation}
and similarly for neutrons.
The numerical values of the matrix elements are \cite{Ellis:2000ds}:
\begin{center}
 $f_{Tu}^{(p)}=0.020 \pm 0.004,\;\; f_{Td}^{(p)}=0.026 \pm 0.005,\;\; f_{Ts}^{(p)}=0.118 \pm 0.062 \; ,$
 \end{center}
 \vspace{-0.5cm}
 \begin{center}
 $f_{Tu}^{(n)}=0.014 \pm 0.003,\;\; f_{Td}^{(n)}=0.036 \pm 0.008$,\;\; $f_{Ts}^{(n)}=0.118 \pm 0.062. \; $ 
 \end{center}
In this model, the elastic scattering of the DM candidate $S_2$ with a nucleon happens via two $t$-channel 
diagrams with $h$ and $H$ as propagators. The resulting spin independent 
scattering cross-section with the assumption of the quark contribution being approximately
equal for both the nucleons (i.e., $f^{(n)}_{Tq}\approx f^{(p)}_{Tq}=f^{N}_{Tq}$) can be written as

\begin{align}\label{eq:cross_sec}
\sigma_N^{SI}=&\frac{\mu^2}{4\pi m_{S_2}^2}\bigg[\frac{\lambda_{HS_2S_2}}{m_H^2}\bigg(\sum_{q=u,d,s}\frac{m_q}{v_0}\cos\alpha\frac{m_N}{m_q}f^N_{Tq}+\frac{2}{27}f^N_{TG}\sum_{q=c,b,t}\frac{m_q}{v_0}\cos\alpha\frac{m_N}{m_q}\bigg)\nonumber\\&
-\frac{\lambda_{hS_2S_2}}{m_h^2}\bigg(\sum_{q=u,d,s}\frac{m_q}{v_0}\sin\alpha\frac{m_N}{m_q}f^N_{Tq}+\frac{2}{27}f^N_{TG}\sum_{q=c,b,t}\frac{m_q}{v_0}\sin\alpha\frac{m_N}{m_q}\bigg)\bigg]^2,
\end{align}
where the reduced mass $\mu=\frac{m_Nm_{S_2}}{m_N+m_{S_2}}$. This formula is an extension of the expression corresponding to the singlet scalar dark matter case \cite{Cline:2013gha}. The relative negative sign between the $h$ and $H$ contibution arises because of the nature of their Yukawa couplings (see Table~\ref{tab:couplings_Yukawa}). Given the presence of the two different channels, we can accommodate the direct detection constraints by appealing to a destructive interference between these two channels (see also Ref.~\cite{Gross:2017dan}).

\begin{figure}[t!]
\includegraphics[scale=0.56]{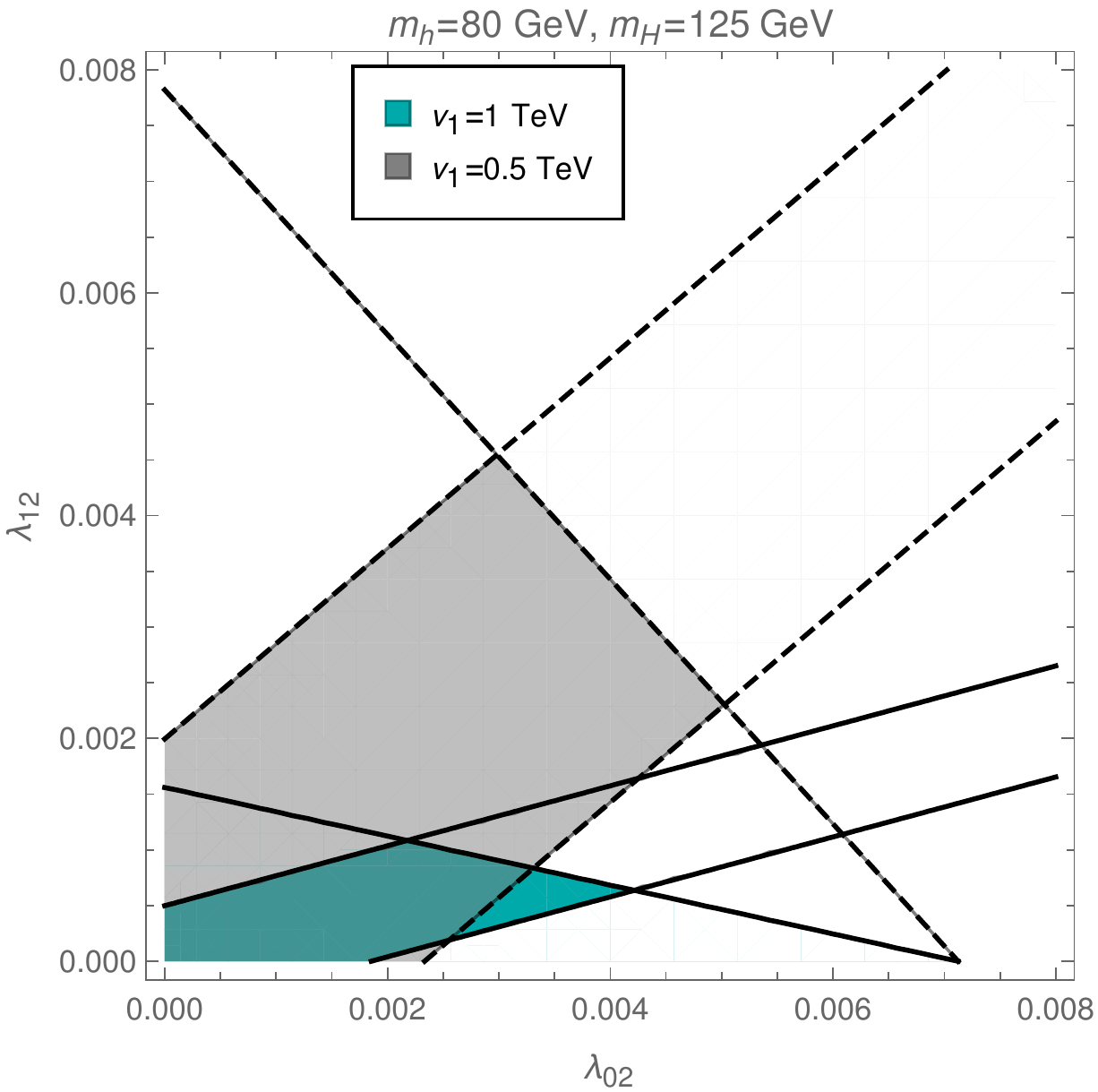}
\includegraphics[scale=0.55]{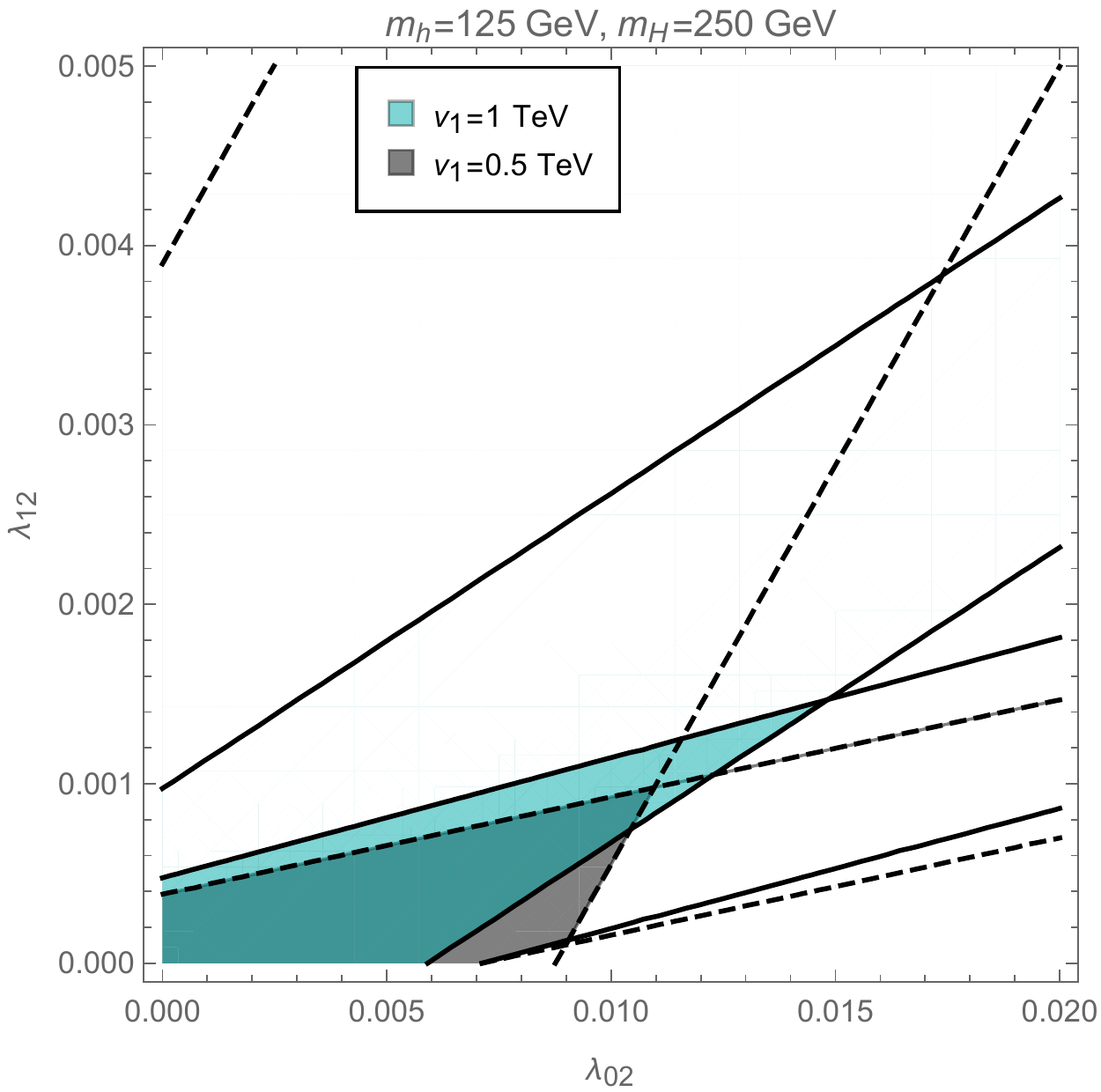}
\caption{The surviving $\lambda_{02}-\lambda_{12}$ parameter space with the twin constraints coming from direct detection (solid lines) and the invisible BR of the higgs (dashed line) imposed for $v_1=1$ TeV (green regions) and $v_1=0.5$ TeV (grey). We have set $\lambda_{01}=0.008$ ($0.04$) for the $m_H=125$ GeV ($m_h=125$ GeV) case.}
\label{fig:direct2}
\end{figure}
 \FloatBarrier
 
The present limit from the Xenon1T result restricts a DM particle of mass 40 GeV (30 GeV) to have a spin independent cross-section lower than  $\sim4.5\times10^{-47}$ $\textrm{cm}^2$($\sim$ $4.1\times10^{-47}$ $\textrm{cm}^2$) \cite{Aprile:2018dbl}. Thus, direct detection bounds put very stringent constraints on $\lambda_{02}$ and $\lambda_{12}$ (since the $hS_2S_2$ and $HS_2S_2$ couplings depend directly on these two parameters). However, $\lambda_{01}$ is not bounded by direct detection directly since the dependence on $\lambda_{01}$ comes only through $\alpha$.
In Fig.~\ref{fig:direct2}, we have combined the bounds on the parameter space of $\lambda_{02}$ and $\lambda_{12}$ from the BR of invisible higgs decay and the direct detection limits together for a 40 GeV $S_2$\footnote{Note that in the one-singlet extension of the SM, a dark matter mass below $\simeq 60$ GeV is not feasible \cite{Guo:2010hq} - the choice here is deliberately one in this window to emphasize the difference in this model.} for two different choices of $v_1$ and for the lowest permissible $\lambda_{01}$ for the two cases $m_h=125$ GeV and $m_H=125$ GeV. In the plots, the solid lines show the allowed range of  $\lambda_{02}$ and $\lambda_{12}$ while taking into account the constraint on $\sigma^{SI}$ and the dashed lines, on the other hand shows the bound on both the couplings complying with the constraint on the BR of invisible higgs decay (as given in Fig.~\ref{fig:invisible}). The shaded region in the intersection of these two constraints is the allowed parameter space in this model - specifically, it can be seen that $\lambda_{12}$ ($\lambda_{02}$) up to 1.1$\times 10^{-3}$ (4.5$\times 10^{-3}$) can be probed for the $m_H=125$ GeV case for $v_1=1$ TeV.

In Table \ref{tab:benchmark}, we indicate a couple of representative benchmark values that can be derived from Fig.~\ref{fig:direct2} for $m_{S_2}=40$ GeV. As expected,\footnote{In Fig.~\ref{fig:direct2}, the two solid lines represent the upper bound of $\sigma^{SI}_N$ and the regions toward the center of the shaded portions accommodate lower values of $\sigma^{SI}_N$.} benchmark points near the edge (of the solid lines) of the contour (BP-1 \& BP-3) provide relatively higher values of $\sigma^{SI}_N$ and benchmark points from a more central region (i.e., equidistant points from the two straight lines of the contour like in BP-2 \& BP-4) can significantly lower the cross-section as evidenced in Table \ref{tab:benchmark}. We will use these benchmark points for further analysis in the subsequent sections.

\begin{table}[h!]
\begin{center}
\begin{tabular}{|c|c|c|c|c|c|c|}
\hline
&Masses (GeV)&$v_1$(GeV)&$\lambda_{01}$ &  $\lambda_{02}$ & $\lambda_{12}$ &  $\sigma^{SI}_N(\textrm{cm}^2)$\\
\hline
BP-1&$m_h=80$, $m_H=125$ &1000&0.008 &  0.002 & 0.0010 & $3.86\times10^{-47}$\\
\hline
BP-2&$m_h=80$, $m_H=125$&1000&0.008 &  0.002 & 0.0005 & $2.61\times10^{-49}$\\
\hline
BP-3&$m_h=125$, $m_H=250$&1000&0.04 &  0.010 & 0.001 & $1.98\times10^{-47}$\\
\hline
BP-4&$m_h=125$, $m_H=250$&1000&0.04 &  0.005 & 0.0006 & $2.35\times10^{-48}$\\
\hline
\end{tabular}
\caption{Representative benchmark points shown along with the corresponding spin independent cross-section values they admit for a dark matter mass of 40 GeV. It is seen that this model offers scope to admit values allowed by the latest Xenon-1T data.}
\label{tab:benchmark}
\end{center}
\end{table}

\begin{figure}[t!]
\includegraphics[scale=0.47]{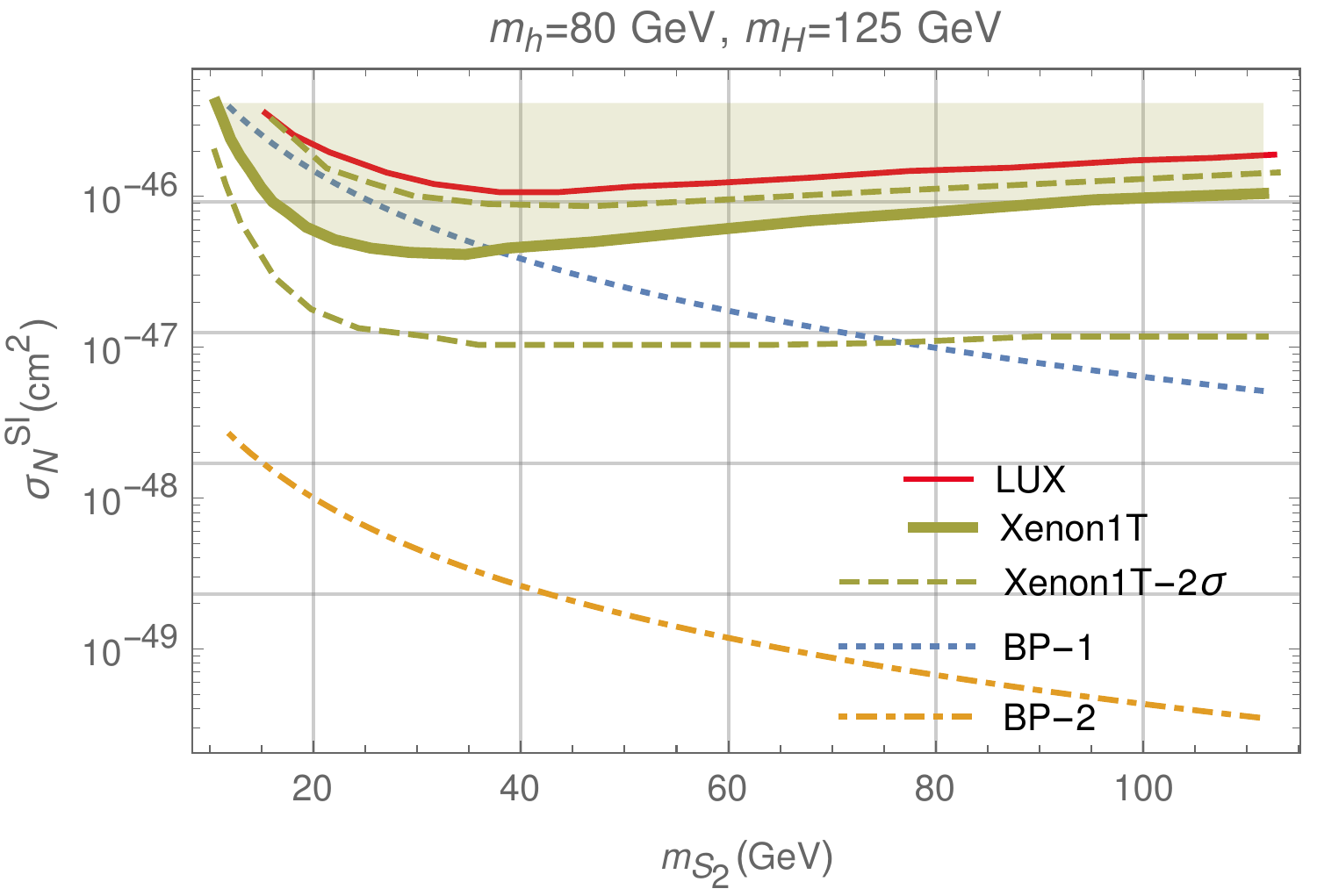}
\includegraphics[scale=0.49]{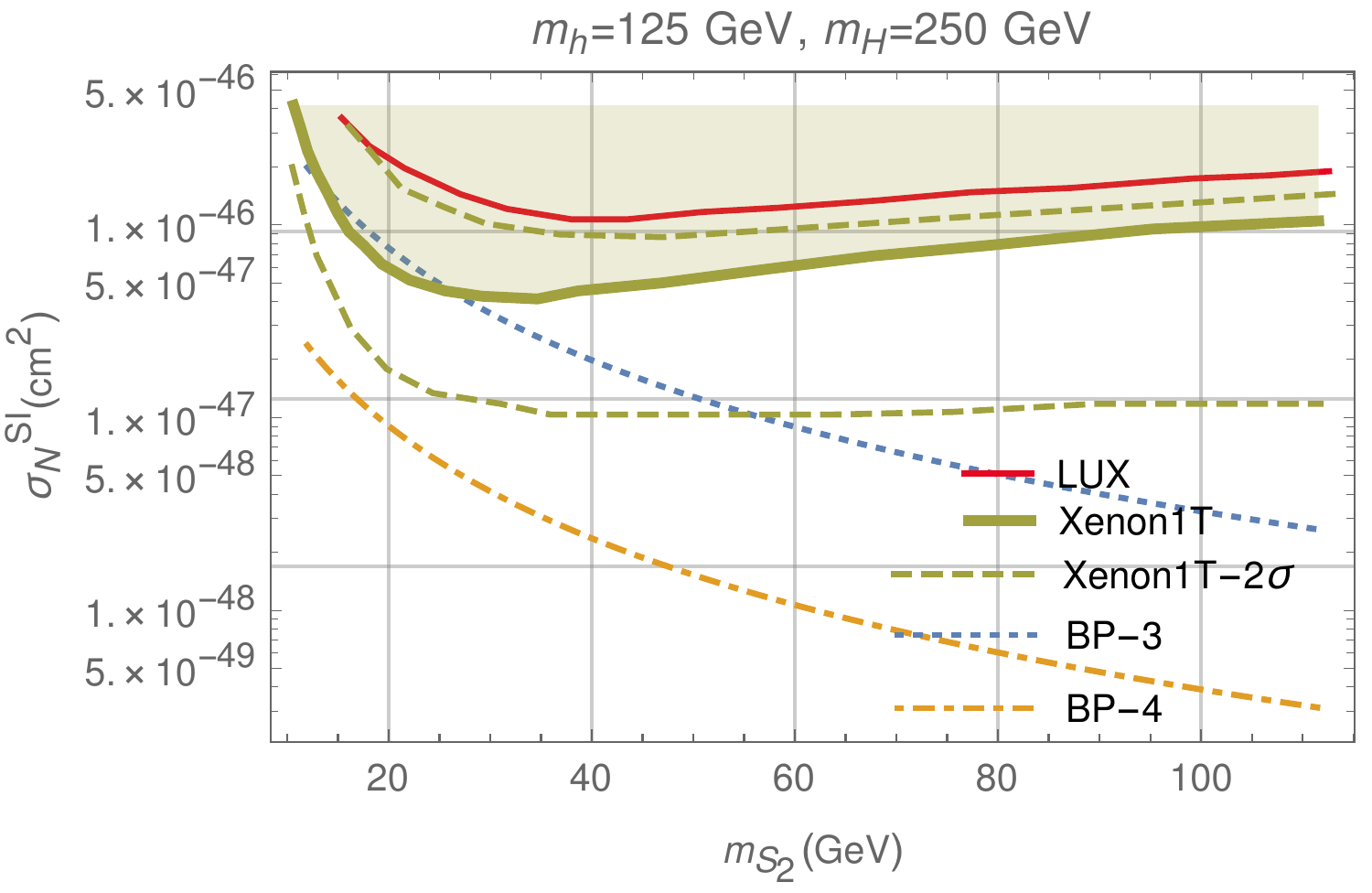}
\caption{Variation of $\sigma^{SI}_N$($\textrm{cm}^2$) with $m_{S_2}$(GeV) for the benchmark points in Table \ref{tab:benchmark} - overlaid on the plots are the constraints from the Xenon-1T and LUX results.}
\label{fig:direct3}
\end{figure}

 In Fig.~\ref{fig:direct3} below, we have shown, for the two sets of benchmark points from Table \ref{tab:benchmark}, 
 the variation of the spin-independent cross-section with the DM mass\footnote{Note that the cross-section values quoted 
 in Table~\ref{tab:benchmark} are for $m_{S_2}=40$ GeV, while in Fig.~\ref{fig:direct3}, we choose the same benchmark points as far as the couplings are considered and vary the DM mass.} - overlaid on the plot are the exclusion limits of Xenon1T \cite{Aprile:2018dbl} and LUX \cite{Akerib:2016vxi}. It can be seen that while BP2 and BP4 can conveniently evade the bound for a wide range of DM masses, BP1 and BP3 cannot do so for lighter $S_2$ - this is so because as these values of couplings aid higher cross-sections, $m_{S_2}$ has to be correspondingly a little higher to bring down the value as $\sigma^{SI}_N\propto m^{-2}_{S_2}$ (see Eqn.~\ref{eq:cross_sec}).


\subsection{Relic Abundance}
\label{sec:relic}

\begin{figure}[t!]
\includegraphics[scale=0.57]{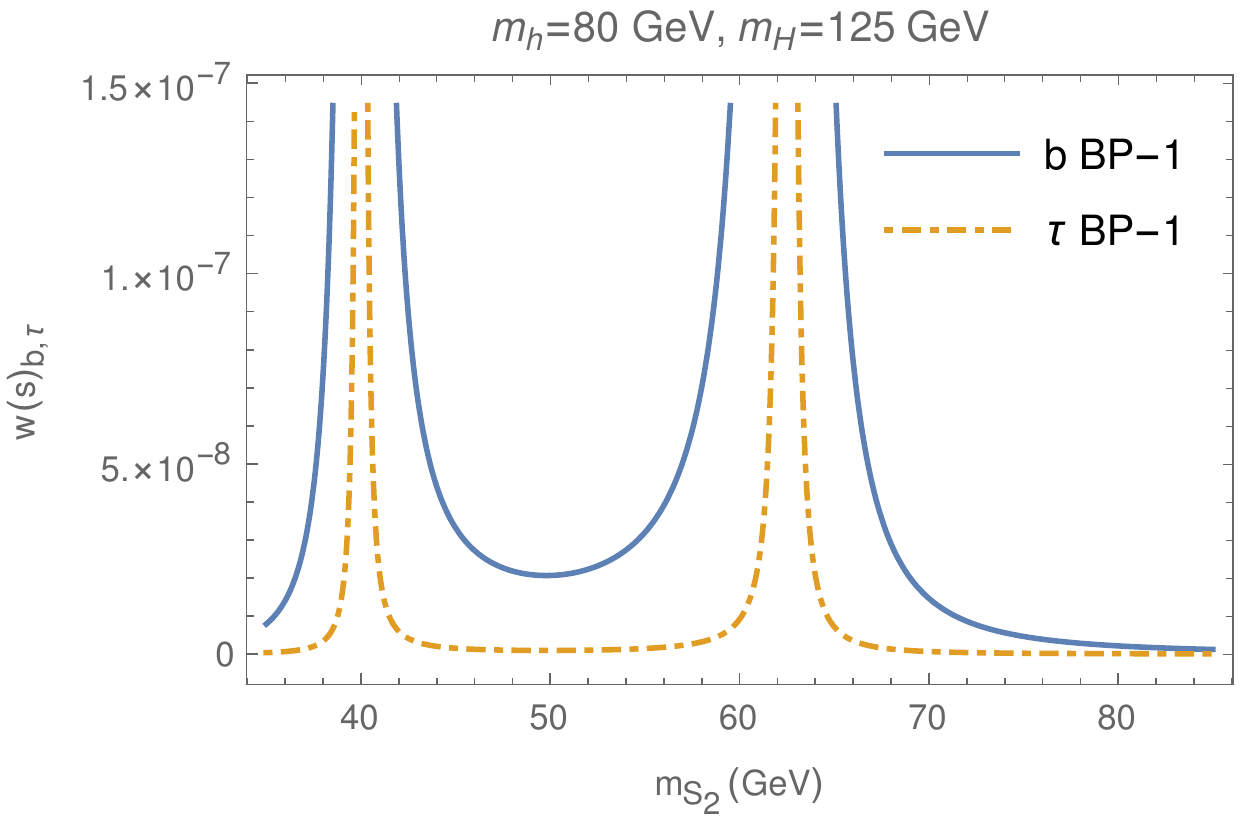}
\includegraphics[scale=0.56]{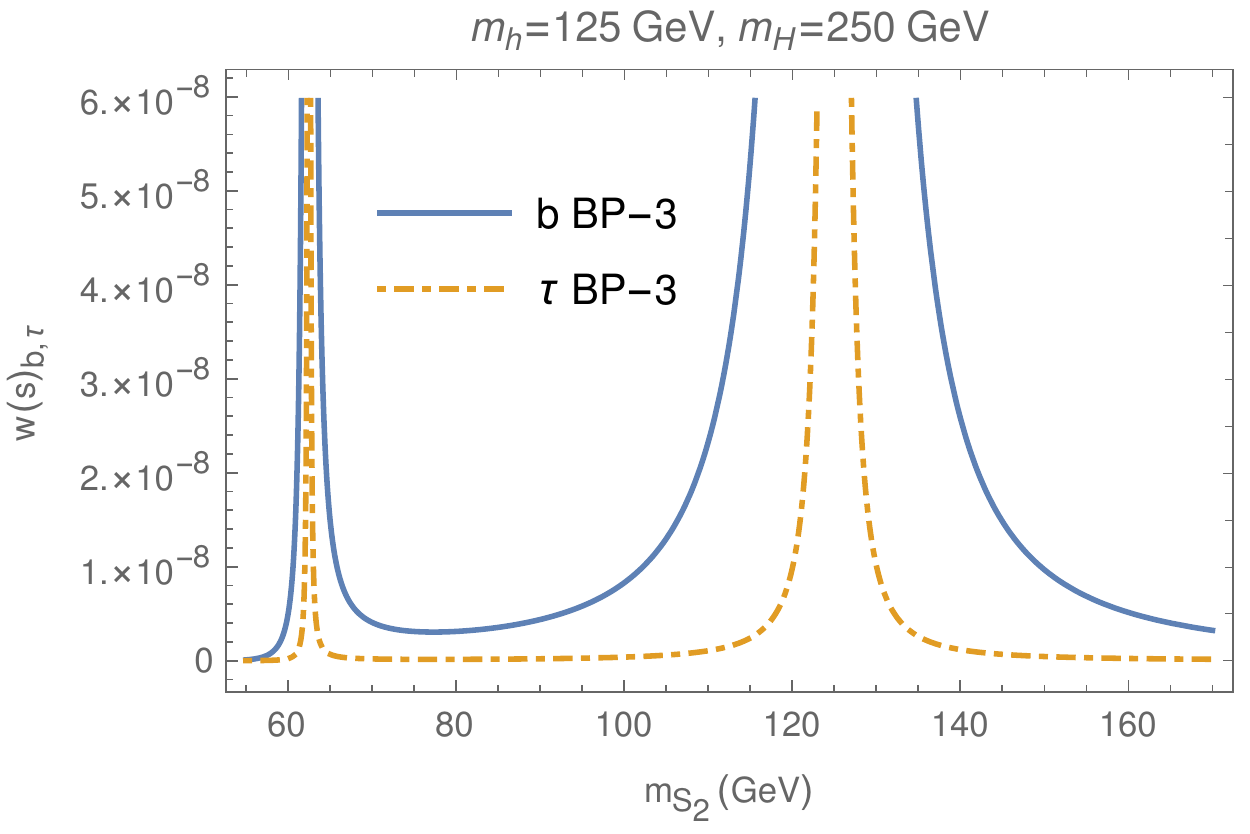}
\caption{Plot of $w(s)$ vs. $m_{S_2}$ showing the contribution of $b\overline{b}$ and $\tau\overline{\tau}$ channels for the benchmark points in Table \ref{tab:benchmark}. The two resonances correspond to $m_h=80$ GeV and $m_H=125$ GeV(left) and $m_h=125$ GeV and $m_H=250$ GeV(right).}
\label{fig:relic1}
\end{figure}

In the early universe, temperature was very high and all the particles including the DM were in thermal equilibrium in the cosmic soup. But as the universe cooled down due to expansion and the interaction rate went 
below the Hubble expansion rate, the annihilation rate of the DM dropped and thus the number density of the DM froze to a constant. This remaining \emph{relic abundance} - now measured to be  $\Omega h^2=0.120\pm0.001$ \cite{planck2018} - is given by  the expression \cite{Kolb_Turner}
\begin{equation}\label{eq:relic}
\Omega h^2=1.07\times 10^9\frac{x_f}{\sqrt{g^*} m_{pl}\langle\sigma v\rangle},
\end{equation}
where $x_f=\frac{m_{S_2}}{T}$, $m_{pl}$ is the Planck mass, and {$g^*$ is the effective degrees of freedom. The velocity averaged cross section $\langle\sigma v\rangle$ up to second order is given by \cite{Srednicki:1988ce}
\begin{equation}\label{eq:sigmav}
\begin{split}
\langle\sigma v\rangle=&\frac{1}{m_{S_2}^2}\bigg[w(s)-\frac{3}{2}\bigg(2w(s)-4m_{S_2}^2w^\prime(s)\bigg)\frac{1}{x_f}\\
&+\frac{3}{8}\bigg(16w(s)-32m_{S_2}^2w^\prime(s)+5(4m_{S_2}^2)^2w^{\prime\prime}(s)\bigg)\frac{1}{x_f^2}\bigg]\bigg|_{s=(2 m_{S_2})^2},
\end{split}
\end{equation}
with $w^\prime(s)=\frac{dw(s)}{ds}$ and
\cite{Srednicki:1988ce,Basak:2013cga}
\begin{equation}\label{eq:w(s)}
w(s)=\frac{1}{32\pi}\sqrt{\frac{s-4m_{S_2}^2}{s}}\int\frac{d(\cos\phi)}{2}\sum_{all\ channels}\lvert \mathscr{M}\rvert^2,
\end{equation}
where $s$ is the center of mass energy.

In this model, the DM annihilates to SM particles only through scalar portal interactions. The relevant channels to consider here are $s$-channel diagrams with $h$ and $H$ as propagators and $b\bar{b}$, 
$\tau\bar{\tau}$, $W^+W^-$, $ZZ$, $hH$, $hh$ and $HH$ as final states. However, given the DM mass we consider here (40 GeV), the gauge and higgs channels will all be highly off-shell and thus the dominant contributions to the annihilation cross section will arise from the $b\bar{b}$ and $\tau\bar{\tau}$ channels.  In Fig.~\ref{fig:relic1}, we have display $w(s)$ as a function of DM mass  depicting the contribution of $b\overline{b}$ (solid line) and $\tau\overline{\tau}$ (dashed line) channels in the annihilation cross section for two of the representative benchmark points listed in Table \ref{tab:benchmark}\footnote{The other two benchmark points also show similar behavior and hence are not displayed here.} - the plot clearly shows the effect of the resonances at $m_h/2$ and $m_H/2$. We observe that away from resonance, the $b\bar{b}$ channel dominates the $\tau\bar{\tau}$ for much of the parameter space.

\begin{figure}[t!]
\includegraphics[scale=0.57]{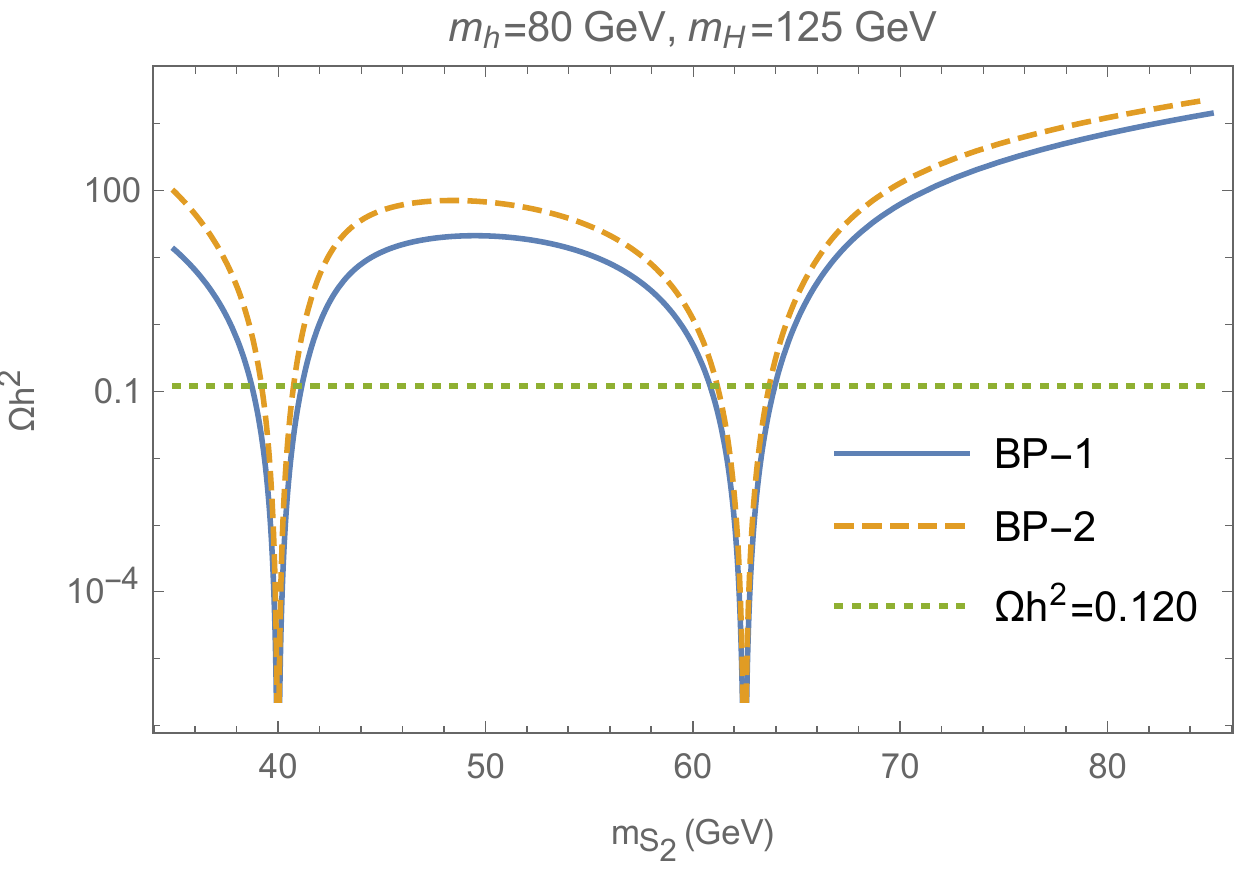}
\includegraphics[scale=0.57]{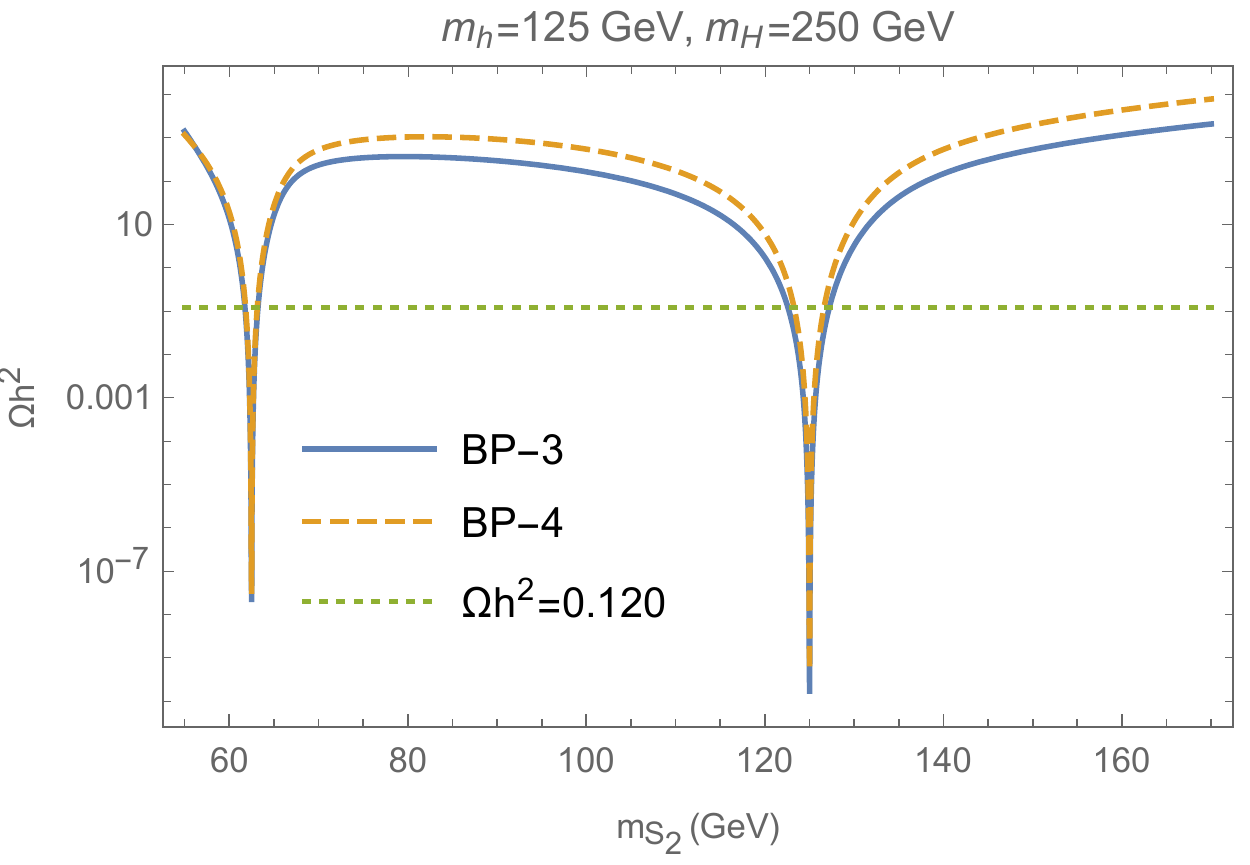}
\caption{Plot of relic abundance as a function of the DM mass for the benchmark points listed in Table \ref{tab:benchmark}. The horizontal dotted line shows the measured value of $\Omega h^2=0.120\pm0.001$.}
\label{fig:relic2}
\end{figure}

In Fig.~\ref{fig:relic2}, we display the relic abundance as a function of the dark matter mass for the set of four benchmark points listed in Table \ref{tab:benchmark}. Due to the Breit-Wigner resonance, an enhancement in the velocity averaged cross section of DM-DM annihilation \cite{guo,murayama} is obtained in certain regions of dark matter mass. Thus, the measured relic abundance can be realized at only the narrow mass regions where the dark matter mass is around half of the mass of either Higgses  $h$ and $H$. This is a general feature of such scalar extension Higgs-portal dark matter models - given the low couplings, one has to resort to a Breit-Wigner enhancement to reproduce the correct relic abundance and hence there is a certain amount of fine-tuning that is necessarily introduced in the model. 

Having consistently explained the dark matter phenomenology in the model consistent with the latest bounds, we now move on to the observed gamma ray excess observed in the Fermi-LAT data and check to see if this simple DM model can account for it.


\subsection{Explaining Gamma-ray excess in Two-singlet model}
\label{sec:gre}

The observed gamma-ray excess found in the Fermi-LAT data points toward a spatially extended excess of $1-3$ GeV gamma rays  \cite{Goodenough:2009gk,Boyarsky:2010dr,Hooper:2010mq,Abazajian:2012pn,Gordon:2013vta} from the regions surrounding the galactic center, the morphology and spectrum of which is best fitted with that predicted from the annihilations of a $36-51$ GeV WIMP into $b\bar b$ with an annihilation 
cross section of $\sigma v = (1-3)\times 10^{-26} \textrm{cm}^3 \textrm{sec}^{-1}$ \cite{Daylan:2014rsa} (normalized to a local dark matter density of $0.4$ $\textrm{GeVcm}^{-3}$). The desired value of the  cross-section is accidentally close to the one required to obtain correct relic abundance. Earlier studies showed that in the simplest one real singlet extension of the SM with a DM  mass in this window fails to explain the gamma-ray excess \cite{Basak:2014sza} as the annihilation cross-section could not be sufficiently enhanced to the desired value. However, in the two singlet extension model again with the aid of the Breit-Wigner enhancement, the required cross section can be enhanced enough to accommodate the gamma ray excess data as shown in Fig.~\ref{fig:gamma} (where we have chosen $m_h=80$ GeV). Given the minimal particle spectrum addition, the two singlet model can thus be regarded as the simplest possible extension of the SM to explain both the dark matter and the gamma ray observation consistently.

\begin{figure}[t!]
\centering
\includegraphics[scale=0.65]{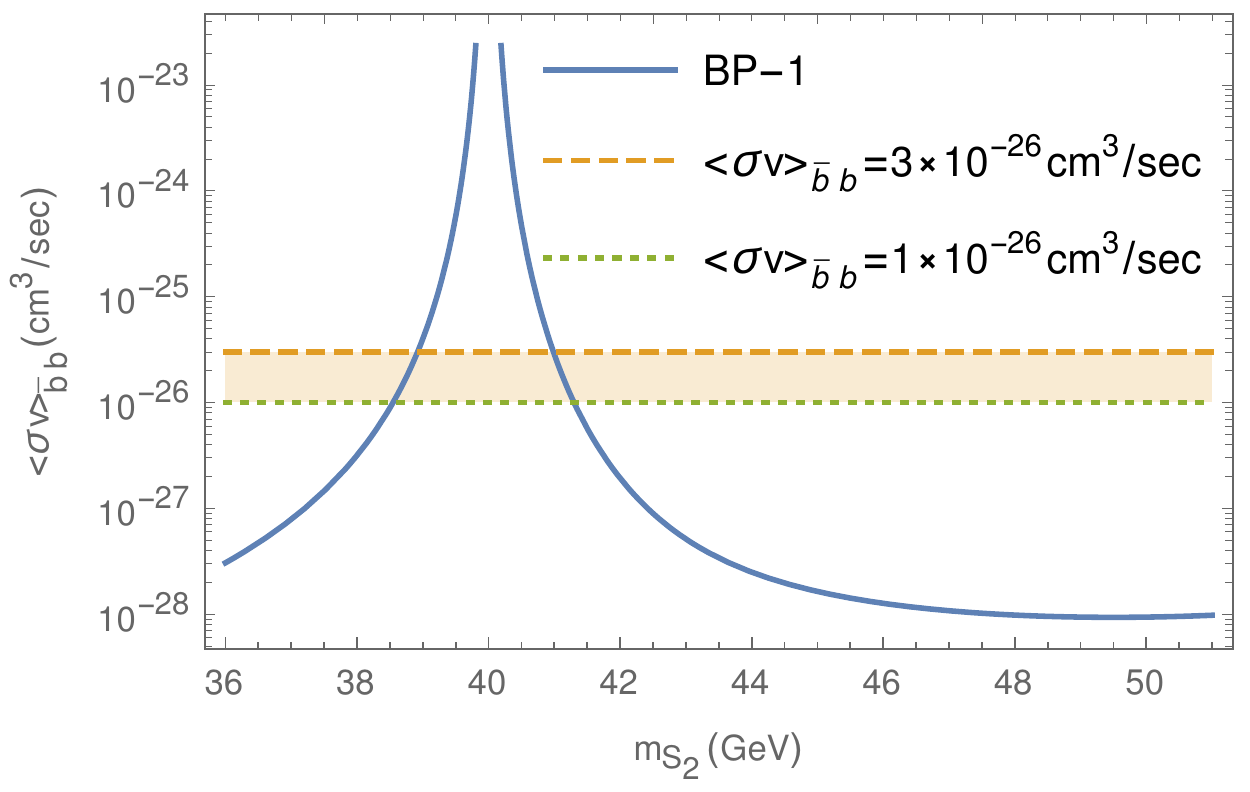} 
\caption{Enhancement in the $\langle\sigma v\rangle_{b\bar{b}}$ values around resonance, shown here as a function of the dark matter mass.}
\label{fig:gamma}
\end{figure}

While the specific mass point $m_{S_2}=40$ GeV certainly works, it is nevertheless interesting to ask if the entire predicted WIMP range of $36-51$ GeV can still accommodate the gamma-ray excess in this model while still complying with all the other constraints put in thus far. Since we will be scanning a slightly wide mass region, constraints from Fig.~\ref{fig:vs1} also have to be modified accordingly - specifically, a  lowering of the $\lambda_{01}$ value is necessary to keep the analysis consistent with the vacuum stability bounds across the mass region of interest while also being consistent with $BR_{h\rightarrow inv}$ and the Xenon1T bounds. In this respect, we find the value $\lambda_{01}= 0.004$ to be consistent with all the given constraints. In Fig.~\ref{fig:gamma2}, we show the allowed region in the $m_{S_2}-m_h$ plane that can explain the gamma ray excess observations\footnote{While we have continued to use the terms BP1 and BP2 in Fig.~\ref{fig:gamma2}, we remind that reader that this is the same set of values as in Table \ref{tab:benchmark} with the one exception of a slightly smaller value of $\lambda_{01}$.} while respecting all the other bounds. The nature of the plot in Fig.~\ref{fig:gamma2} can be understood by appealing to Fig.~\ref{fig:gamma} - note that there are two sets of mass points that can each accommodate the lower or the higher $\sigma v$ in the range. Specifically, $m_{S_2}\simeq 38.6$ and $41.3$ GeV correspond to $\sigma v = 1\times 10^{-26} \textrm{cm}^3 \textrm{sec}^{-1}$ and $m_{S_2}\simeq 38.9$ and $41$ GeV correspond to $\sigma v = 3\times 10^{-26} \textrm{cm}^3 \textrm{sec}^{-1}$. This same feature carries over in Fig.~\ref{fig:gamma2} - the borders of each thick colored line represents the two ends of the range in $\sigma v = (1-3)\times 10^{-26} \textrm{cm}^3 \textrm{sec}^{-1}$, and the upper and lower colored lines indicate that these ranges can correspond to two different masses. The regions shaded in blue and orange can accommodate the Gamma ray excess while still conforming to all the other constraints. The reliance on the Breit-Wigner resonance thus limits the parameter space to one in which the two masses are highly correlated. Also, since we want an enhancement in the low mass region, the possibility that the lighter of the two Higgses is the observed 125 GeV Higgs cannot be considered and we are forced to choose the $m_h=80$ GeV scenario.

\begin{figure}[t!]
\centering
\includegraphics[scale=0.5]{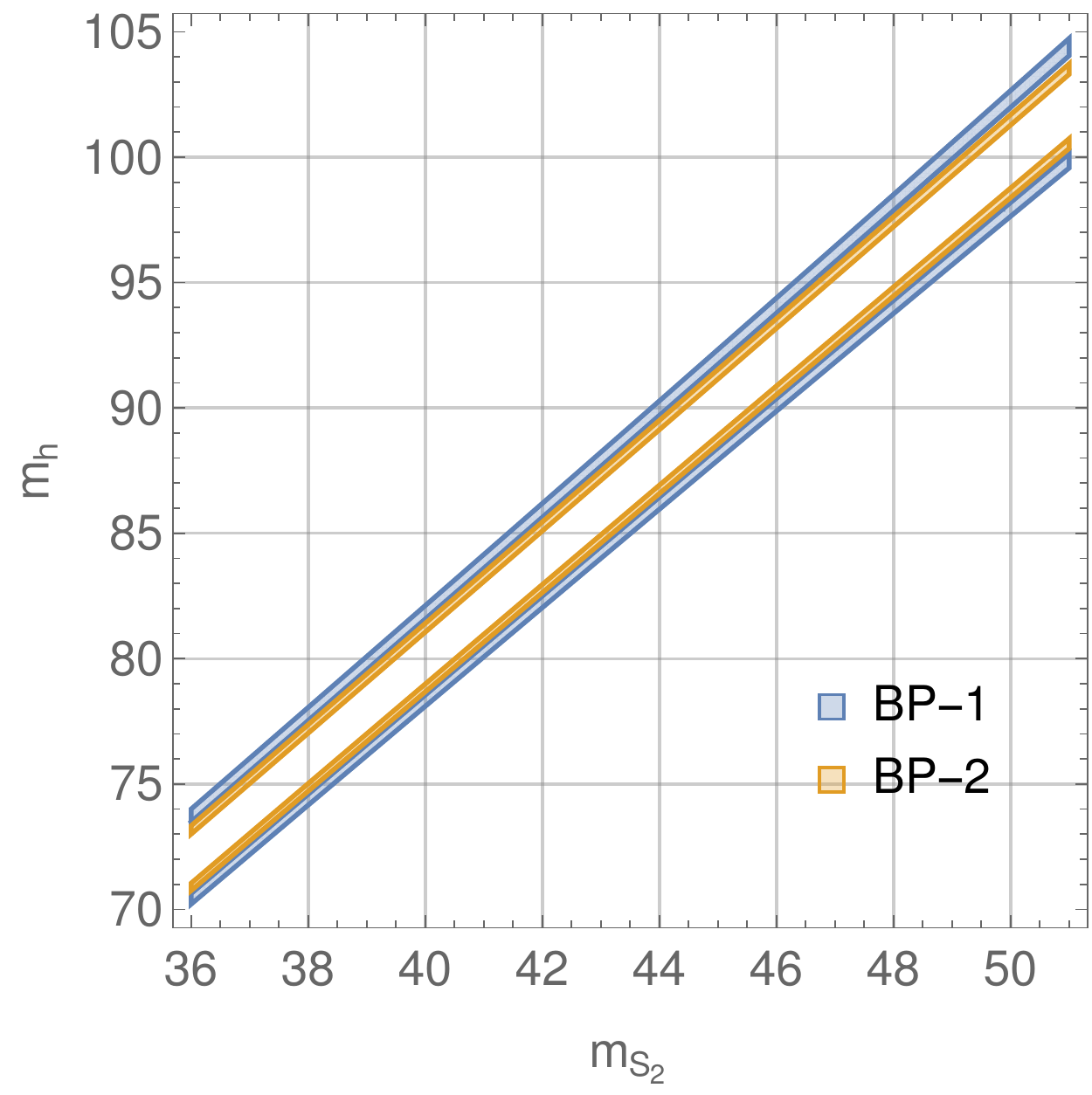}
\caption{The region allowed by the gamma ray excess observation in mass space for $\lambda_{01}=0.004$ and $v_1$, $\lambda_{02}$, $\lambda_{12}$ from Table \ref{tab:benchmark} - the surviving regions is shown in the shaded blue and orange lines for the benchmark points BP1 and BP2.}
\label{fig:gamma2}
\end{figure}


\section{Conclusion}
\label{sec:conclusion}

In this work, we studied a simple scalar extension of the SM with two real singlets $S_1$ and $S_2$ with a $Z_2\times Z_2^\prime$ symmetry. While the $Z_2$ is broken, the other discrete $Z_2^\prime$ remains unbroken enabling the scalar $S_2$ to be a potential dark matter candidate. The interaction strength of the singlet scalars with the SM sector is dictated via a Higgs portal term in the Lagrangian of the form $\lambda_{0i}(\Phi^\dagger\Phi)S_i^2$. The model has the advantage that with a minimal addition to the SM particle spectrum, it provides a dark matter candidate consistent with the latest experimental bounds while also simultaneously accommodating the observed gamma ray excess. 

After putting in the vacuum stability constraints on the Higgs potential, we find that the coupling $\lambda_{01}$ is constrained to lie in the $10^{-1}-10^{-2}$ range (depending on whether the light or the heavier Higgs is associated with the 125 GeV Higgs). While $\lambda_{02}$ is unconstrained by vacuum stability considerations as it does not mix with the other CP-even scalars, direct bounds from the Higgs invisible branching ratios forces $\lambda_{02}\approx 10^{-2}-10^{-3}$ thus severely restricting the couplings of the singlets to the SM sector. Interestingly, even though the direct detection exclusion limits coming from experiments like Xenon1T shrink the coupling parameter space appreciably, the model still has potential discoverable regions. Typically, a model can be ruled out if the upper limit on the couplings imposed by the direct detection experiments are insufficient to generate the required relic density values. However, in the two singlet model, the direct detection constraint is relaxed to a great extent by the destructive interference between the two $t$-channel $h$ and $H$ propagators while the Breit-Wigner resonance helps to get the correct relic abundance. In addition, we have also demonstrated that the two singlet model can also account for the gamma-ray excess for an $S_2$ in the range $36-51$ GeV - however it has to be noted that this explanation imposes the restriction that we cannot consider the possibility of the observed 125 GeV Higgs to be the lighter of the two CP-even eigenstates and are forced to choose $m_h=80$ GeV and $m_H=125$ GeV.

The two singlet extension thus remains an attractive possibility from a dark matter perspective. Future limits from the Higgs invisible branching ratio estimates and the Xenon-1T experiment would undoubtedly constrain the parameter space further. The model is also interesting from a collider physics angle. While the restriction of the coupling strength of the extra scalars to the SM fermions would make direct collider searches of these scalars at the LHC difficult, it nevertheless is interesting to ask if one could uncover these Higgs portal signatures by appealing to unique multi-Higgs final states that have a relatively low SM background - these remain interesting avenues to explore.

\appendix

\section{Relevant Couplings}\label{sec:appendix}
In this Appendix, we list all the relevant couplings of the extra scalars that have been used in this study.

\begin{table}[h!]
\begin{center}
\begin{tabular}{|c|c|}
\hline
$\lambda_{abc}$& Couplings in terms of Lagrangian parameters\\
\hline
$HHH$&$\frac{\lambda_0v_0}{4}\cos^3\alpha+\lambda_1v_1\sin^3\alpha+\lambda_{01}v_0\cos\alpha\sin^2\alpha+\lambda_{01}v_1\sin\alpha\cos^2\alpha$\\
\hline
$hhh$&$-\frac{\lambda_0v_0}{4}\sin^3\alpha+\lambda_1v_1\cos^3\alpha-\lambda_{01}v_0\sin\alpha\cos^2\alpha+\lambda_{01}v_1\cos\alpha\sin^2\alpha$\\
\hline
$HHh$ &$-\frac{3\lambda_0v_0}{4}\cos^2\alpha\sin\alpha+3\lambda_1v_1\sin^2\alpha\cos\alpha$\\
&$+\lambda_{01}v_0(2\sin\alpha\cos^2\alpha-\sin^3\alpha)+\lambda_{01}v_1(\cos^3\alpha-2\sin^2\alpha\cos\alpha)$\\
\hline
$Hhh$ &$\frac{3\lambda_0v_0}{4}\cos\alpha\sin^2\alpha+3\lambda_1v_1\sin\alpha\cos^2\alpha$\\
&$+\lambda_{01}v_0(\cos^3\alpha-2\sin^2\alpha\cos\alpha)+\lambda_{01}v_1(\sin^3\alpha-2\sin\alpha\cos^2\alpha)$\\
\hline
$HS_2S_2$&$\lambda_{02}v_0\cos\alpha+2\lambda_{12}v_1\sin\alpha$\\
\hline
$hS_2S_2$&$-\lambda_{02}v_0\sin\alpha+2\lambda_{12}v_1\cos\alpha$\\
\hline
\end{tabular}
\caption{The cubic couplings of the various Higgs bosons in the two singlet model.}
\label{tab:couplings_3}
\end{center}
\end{table}

\begin{table}[h!]
\begin{center}
\begin{tabular}{|c|c|}
\hline
$\lambda_{abcd}$& Couplings in terms of Lagrangian parameters\\
\hline
$HHHH$&$\frac{\lambda_0}{16}\cos^4\alpha+\frac{\lambda_1}{4}\sin^4\alpha+\frac{\lambda_{01}}{2}\cos^2\alpha\sin^2\alpha$\\
\hline
$hhhh$&$\frac{\lambda_0}{16}\sin^4\alpha+\frac{\lambda_1}{4}\cos^4\alpha+\frac{\lambda_{01}}{2}\cos^2\alpha\sin^2\alpha$\\
\hline
$S_2S_2S_2S_2$&$\frac{\lambda_2}{4}$\\
\hline
$HHhh$&$\frac{3\lambda_0}{8}\sin^2\alpha\cos^2\alpha+\frac{3\lambda_1}{2}\sin^2\alpha\cos^2\alpha+\frac{\lambda_{01}}{2}(\cos^4\alpha+\sin^4\alpha-4\sin^2\alpha\cos^2\alpha)$\\
\hline
$HHS_2S_2$&$\frac{\lambda_{02}}{2}\cos^2\alpha+\lambda_{12}\sin^2\alpha$\\
\hline
$hhS_2S_2$&$\frac{\lambda_{02}}{2}\sin^2\alpha+\lambda_{12}\cos^2\alpha$\\
\hline
$HHHh$&$-\frac{\lambda_0}{4}\sin\alpha\cos^3\alpha+\lambda_1\sin^3\alpha\cos\alpha+\lambda_{01}(\sin\alpha\cos^3\alpha-\sin^3\alpha\cos\alpha)$\\
\hline
$Hhhh$&$-\frac{\lambda_0}{4}\sin^3\alpha\cos\alpha+\lambda_1\sin\alpha\cos^3\alpha+\lambda_{01}(\sin^3\alpha\cos\alpha-\sin\alpha\cos^3\alpha)$\\
\hline
$HhS_2S_2$&$-\lambda_{02}\sin\alpha\cos\alpha+2\lambda_{12}\sin\alpha\cos\alpha$\\
\hline
\end{tabular}
\caption{The quartic scalar couplings of the various Higgs bosons in the two singlet model.}
\label{tab:couplings_4}
\end{center}
\end{table}

\begin{table}[h!]
\begin{center}
\begin{tabular}{|c|c|}
\hline
$\lambda_{abc}$& Couplings \\
\hline
$Hb\bar{b}$&$\frac{-\textrm{e}m_b}{2\sin\theta_\omega m_W}\cos\alpha$\\
\hline
$hb\bar{b}$&$ \frac{\textrm{e}m_b}{2\sin\theta_\omega m_W}\sin\alpha$\\
\hline
$H\tau\bar{\tau}$&$\frac{-\textrm{e}m_\tau}{2\sin\theta_\omega m_W}\cos\alpha$\\
\hline
$h\tau\bar{\tau}$&$ \frac{\textrm{e}m_\tau}{2\sin\theta_\omega m_W}\sin\alpha$\\
\hline
\end{tabular}
\caption{The relevant Yukawa couplings of the $h$ and $H$ with the SM fermions.}
\label{tab:couplings_Yukawa}
\end{center}
\end{table}

\begin{table}[h!]
\begin{center}
\begin{tabular}{|c|c|}
\hline
$\lambda_{abc}$& Couplings \\
\hline
$HWW$&$\frac{em_W\cos\alpha}{\sin\theta_\omega}$\\
\hline
$hWW$&$-\frac{em_W\sin\alpha}{\sin\theta_\omega}$\\
\hline
$HZZ$&$\frac{em_Z\cos\alpha}{2\sin\theta_\omega\cos\theta_\omega}$\\
\hline
$hZZ$&$-\frac{em_Z\sin\alpha}{2\sin\theta_\omega\cos\theta_\omega}$\\
\hline
\end{tabular}
\caption{The relevant scalar-gauge boson couplings in the two singlet model.}
\label{tab:couplings_gauge}
\end{center}
\end{table}
\FloatBarrier
\begin{acknowledgments}
BC acknowledges support from the Department of Science and Technology, India, under Grant CRG/2020/004171.
\end{acknowledgments}

\bibliographystyle{apsrev}
\bibliography{Ref_Paper}

\end{document}